\documentclass[12pt,draftclsnofoot,letterpaper,onecolumn,romanappendices]{IEEEtran}
\usepackage{cite}
\usepackage{epstopdf}
\usepackage{epsfig}
\usepackage{amsmath,amssymb,amsfonts,amstext,amsbsy,amsopn,dsfont}
\usepackage{cases}
\usepackage{bigints}
\usepackage{sublabel}
\usepackage{array}

\newtheorem{theorem}{\textbf{Theorem}}

\newtheorem{lemma}{\textbf{Lemma}}

\newcommand{\defn}{\triangleq}
\newcommand{\dif}{\textmd{d}}
\newcommand{\Lt}{\mathsf{L}}
\newcommand{\Ut}{\mathsf{U}}

\newcommand{\X}{\mathcal{X}}
\newcommand{\U}{\mathcal{U}}

\begin{document}

\title{Traffic Management for Heterogeneous Networks with Opportunistic Unlicensed Spectrum Sharing}

\author{Chun-Hung Liu and Hong-Cheng Tsai
\thanks{C.-H. Liu and H.-C. Tsai are with the Dept. of Electrical and Computer Engineering, National Chiao Tung University, Hsinchu, Taiwan. The contact author is Dr. Liu. (Email: chungliu@nctu.edu.tw)  Manuscript date: \today.}}

\maketitle

\begin{abstract}
This paper studies how to maximize the per-user-based throughput in an $M$-tier heterogeneous wireless network (HetNet) by optimally managing traffic flows among the access points (APs) in the HetNet. The APs in the first $M-1$ tiers can use the licensed spectrum at the same time whereas they share the unlicensed spectrum with the APs in the $M$th tier by the proposed opportunistic CSMA/CA protocol. We characterize the statistical property of the cell load and channel access probability of each AP using a general AP association scheme. For an AP in each tier, the tight bounds on its mean spectrum efficiencies in the licensed and unlicensed spectra are derived in a low-complexity form for general random channel gain and AP association weight models and they can give some insights on how channel gains, AP association weights and void AP probabilities affect the mean spectrum efficiencies. We define the per-user link throughput and per-user network throughput based on the derived the mean spectrum efficiencies and maximize them by proposing the decentralized and centralized traffic management schemes for the APs in the first $M-1$ tiers under the constraint that  the per-user link throughput of the tier-$M$ APs must be above some minimum required value. Finally, a numerical example of coexisting LTE and WiFi networks is provided to validate our derived results and findings.
\end{abstract}


\section{Introduction}
As more and more versatile services are offered over wireless networks and new generations of wireless smart handsets get wider and wider adoption, considerable data traffic flowing over spectrum-limited cellular networks is an inevitable phenomenon the network operators have to seriously face. To alleviate the spectrum crunch crisis of a cellular network, an effective means is to make a traditional cellular network migrate to a heterogeneous cellular network in which
many different kinds of base stations (BSs), such as macrocell, micro and small cell BSs, are densely deployed. Although heterogeneous cellular networks have a much higher network capacity compared to their traditional counterparts, their licensed spectrum is still very limited and their per-user link throughput may not be efficiently improved if the network has a huge user population. Accordingly, exploiting more available spectrum for heterogeneous cellular networks is the right track that should be followed, which fosters the idea of extending the service of the cellular BSs to the unlicensed spectrum. 

If cellular BSs can access the licensed and unlicensed spectra at the same time, they can integrate all available spectrum resources by using the carrier aggregation technique\cite{MBMSACO13,HZXCWGSW15,RZMWLXCZZXSLLX15}. However, extending cellular services to the unlicensed spectrum could severely impact the throughput performance of the existing access points (APs) using the unlicensed spectrum, such as WiFi APs. This coexisting problem in the unlicensed spectrum motivates us to investigate how to make different kinds of BSs and APs properly share the unlicensed spectrum and improve their total throughput is an important problem that needs to be investigated thoroughly. To generally and tractably analyze the throughput performance of a heterogeneous wireless network (HetNet) with unlicensed spectrum sharing, the HetNet considered in this paper has $M$ tiers. All APs in the same tier of the HetNet are of the same type and performance and they follow an independent Poisson point process (PPP) with certain intensity. Specifically, the APs in the $M$th tier only access the unlicensed spectrum whereas the APs in the first $M-1$ tiers can simultaneously access the licensed spectrum as well as the unlicensed spectrum if they have a chance. All APs use the proposed (slotted nonpersistant) opportunistic CSMA/CA with random backoff time protocol to contend the channel in the unlicensed spectrum. Such a HetNet model characterizes the coexisting impacts among different kinds of APs in the unlicensed spectrum while the CSMA/CA protocol is adopted.  

\subsection{Prior Works on Unlicensed Spectrum Sharing}
Earlier studies on the coexisting interference problem in the unlicensed spectrum focused on how to make APs in different overlaid wireless networks share the unlicensed spectrum with certain fairness. For example, reference \cite{REAPDT07} proposed a game-theoretical approach to fairly sharing the unlicensed spectrum in multiple coexisting and interfering networks. The interference modeling and mitigation problems in the unlicensed spectrum are investigated in references  \cite{HYPPHCNRP07,YMSKAH09,JBENNJR12}. These works use a small-scale and deterministic network model to formulate their problems and they do not study some coexisting performance metrics, such as coverage and network throughput. Although the authors in \cite{JJQLHNAPGW14} developed a more accurate interference analysis for large-scale networks based on the continuum field approximation; however, they did not investigate if the interference in the unlicensed spectrum can be effectively mitigated by using channel access protocols. 

A more complicate coexistence problem in the unlicensed spectrum that recently attracts a lot of attentions is how to let the base stations (BSs) or APs originally use the licensed spectrum also be able to access the unlicensed spectrum and use the carrier aggregation technique to boost their overall throughput\cite{HZXCWGSW15,RZMWLXCZZXSLLX15}. A few recent works have already shown that LTE and WiFi networks coexisting in the unlicensed spectrum can significantly improve their entire network throughput\cite{HZXCWGSW15,RZMWLXCZZXSLLX15,CHLHCT16,XDCHLLCWXZ16,HCTCHL16}. However, how to manage the traffic flows between LTE BSs and WiFi APs to maximize the total or per-user link throughput in the licensed and unlicensed spectra is not addressed in these works.  Although a recent work in \cite{QCGYHSAMGYLAH16} studied when to offload the traffic from the LTE network to the WiFi network and share the unlicensed spectrum in order to maximize the per-user link throughput, where their network, interference and spectrum sharing models are too much simple so that their observations may not be applicable to a large-scale stochastic network. Offloading traffic from an LTE network to another WiFi network may not  increase the per-user link throughput since the offloaded users lose their licensed throughput. Hence, as long as the LTE and WiFi networks can coexist without causing severe interference, having them sharing the unlicensed spectrum is a good policy. In \cite{AREWPCAIZD15,FLEBEEMCBRY15}, small cell BSs are shown to achieve a notable throughput gain if they can cleverly and properly access the unlicensed spectrum without causing much interference to the WiFi APs. 

A stochastic-geometry-based framework in \cite{ABCIPZ14,SSISDR15,XDCHLLCWXZ16,CHLLCW1602,YLFBJGA16} is applied to analyze the coexistence performance of large-scale LTE and WiFi networks, but the exact analyses of the coexisting throughputs of the APs in the licensed and unlicensed spectra were not studied in these works. Although few existing works indeed analyzed the throughput problem in coexisting LTE and WiFi networks, their throughput analysis is too simple to evaluate the link/network throughput very accurately. For example, 
reference \cite{XWTQMSJL17} studied the network throughput by using the minimum required signal-to-interference power ratio (SIR) of each link and this network throughput may not be close to the real network throughput in that the link throughput of each AP could be significantly underestimated. In addition, these aforementioned prior works did not use a general network model to theoretically analyze the link throughput of each AP while the unlicensed spectrum is shared so that they are unable to offer some insights on how to maximize the per-user link throughput of each AP and overall network throughput by managing the traffic among all APs.

\subsection{Contributions}
The main contributions in this paper are summarized in the following:
\begin{itemize}
	\item \textit{A general $M$-tier HetNet architecture is proposed}: This $M$-tier network model is more general than the models in the aforementioned prior works so that it is able to generally characterize the licensed and unlicensed spectrum sharing problem among different kinds of APs using the opportunistic CSMA/CA protocol. Under this HetNet model, the coexisting interference models in the licensed and unlicensed spectra can be easily proposed.
	\item\textit{A general AP association scheme is proposed}: It is able to cover several pathloss-based AP association schemes is adopted in the HetNet.
	\item\textit{The void probability and association probability for an AP in each tier are accurately found based on the proposed AP association scheme}: We show that The void probability of densely-deployed APs is in general not small and thus cannot be ignored. However, many previous works overlooked this important fact and thus their analytical results could be very inaccurate.
	\item \textit{The integral identity of the Shannon transformation is derived}: This identity can be applied to find the compact result of the mean spectrum efficiency of a wireless link in a Poisson wireless network.
	\item \textit{The decentralized and centralized traffic management schemes are proposed}: The decentralized traffic management is able to maximize the per-user link throguhput of each AP with limited local information, whereas the the centralized traffic management scheme can maximize the per-user network throughput if the information of all APs can be processed jointly.
\end{itemize}

In this paper, we are only able to find the tight lower bounds on the mean spectrum efficiencies of the APs in each tier since the resulting transmitting APs in each tier are no longer a PPP due to performing AP association and CSMA to access the unlicensed channel. Nonetheless, with the aid of the integral identity of the Shannon transform, these tight lower bounds are derived without assuming any specific channel gain and AP association weight models. Most importantly, they are low-complexity so that they can give us some intuitions regarding how they are affected by channel gain impairments, AP association weights as well as void AP probabilities. This feature is absolutely important since we can easily judge whether the mean spectrum efficiency of the APs in a particular tier increases or decreases due to traffic loading or offloading of the APs in other tiers.
 
For the  strategy of traffic management in this paper, our idea is to make the APs in the first $M-1$ tiers that can access the licensed and unlicensed spectra achieve the per-user throughput as much as possible, whereas the APs in the $M$th tier that only access the unlicensed spectrum just need to have their per-user link throughput above some minimum required value. The \textit{per-user link throughput} is defined based on the spectrum efficiencies of the APs in the licensed and unlicensed spectra. We propose a decentralized traffic manage scheme that maximizes the per-user link throughput of the APs in the first $M-1$ tiers and maintains the per-user link throughput requirement of the APs in the $M$th tier. We also propose a centralized traffic management scheme that can be performed by the central processing unit of the core network of the HetNet to maximize the defined per-user network throughput by managing traffic offloading or loading of all APs. Moreover, the network model in this paper is built based on the assumption that all APs in each tier follows a homogeneous PPP so that the locations of all deployed APs are completely independent. Thus, the analytical results found in this paper can be used to do the worse-case performance evaluations for the LTE-U (LTE in Unlicensed spectrum) or LAA (Licensed-Assisted Access) system in 3GPP since the APs in a real LTE-U/LAA system may be deployed with location correlation\cite{3GPPLAA15}. Finally, a numerical simulation example of coexisting LTE BSs and WiFi APs is provided and the numerical results validate the accuracy and correctness of our analytical results and findings.

\section{System Model and Preliminaries}\label{Sec:SystemModel}
Consider a large-scale interference-limited HetNet consisting of $M$ tiers of access points (APs). All the APs in the same tier are of the same type and performance. Specifically, the APs in the $m$-th tier, denoted by set $\X_m$, follow an independent marked Poisson point process (PPP) of intensity $\lambda_m$ defined as follows
\begin{align}
\X_m\defn\{(X_{m_i},P_m,V_{m_i}): X_{m_i}\in\mathbb{R}^2, P_m\in\mathbb{R}_{+},V_{m_i}\in\{0,1\}, i\in\mathbb{N}_+\},\, m\in\mathcal{M},
\end{align}
where $\mathcal{M}\defn\{1,2,\ldots,M\}$ , $X_{m_i}$ denotes AP $i$ in the $m$-th tier and its location, $P_m$ is the transmit power used by the APs in the $m$-th tier, and $V_{m_i}$ is a Bernoulli random variable indicating whether AP $X_{m_i}$ is void or not: if AP $X_{m_i}$ is associated with at least one user (i.e., it is not void), then $V_{m_i}=1$ and zero otherwise. Without loss of generality, we assume the APs in the $M$th tier only use the unlicensed spectrum to deliver data, and all other APs in the first $M-1$ tiers primarily use the licensed spectrum and opportunistically use the unlicensed spectrum by carrier aggregation to transmit data if they have a chance to access the unlicensed spectrum. This network model with unlicensed spectrum sharing has a practical application context. For example, in a heterogeneous cellular network, LTE-U macrocell and small cell base stations (BSs) consisting of the APs in the first $M-1$ tiers can coexist and share the unlicensed spectrum with WiFi APs in the $M$th tier if the LTE-U BSs can use the carrier aggregation technique to integrate the licensed and unlicensed spectrum resources \cite{RZMWLXCZZXSLLX15,HZXCWGSW15}. 

All users also follow an independent PPP $\U$ of intensity $\mu$ given by
\begin{align}
\U\defn\{U_j:U_j\in\mathbb{R}^2, \forall j\in\mathbb{N}_+\}
\end{align}
and we assume there is typical user $U_0$ located at the origin without loss of generality. Our following location-dependent analyses will be based on typical user $U_0$ for simplicity since the analytical results do not depend where the typical user is located due to Slivnyak's theorem\cite{DSWKJM96}. We consider a \textit{downlink} transmission scenario in this paper and each user selects its serving AP $X_o$ by adopting the following AP association scheme 
\begin{align}\label{Eqn:APAssoScheme}
X_o\defn \arg \sup_{X_{m_i}\in\bigcup_{m=1}^M\X_m} W_{m_i}\|X_{m_i}\|^{-\alpha}=\arg\inf_{X_{m_i}\in\bigcup_{m=1}^M\X_m} W^{\frac{1}{\alpha}}_{m_i}\|X_{m_i}\|,
\end{align}
where $W_{m_i}$ is the random AP association weight with mean $\bar{w}_m$ for AP $X_{m_i}$, $\|X_i-X_j\|$ denotes the distance between nodes $X_i$ and $X_j$ for $i\neq j$, and $\|X_{m_i}\|^{-\alpha}$ is called the pathloss of AP $X_{m_i}$ with pathloss exponent $\alpha>2$. Furthermore, we assume that all $W_{m_i}$'s are independent, all $\frac{W_{m_i}}{\bar{w}_m}$'s are i.i.d. random variables with unit mean, and the $a$-fractional moment of $W_{m_i}$ always exists for all $i\in\mathbb{N}_+$ and $m\in\mathcal{M}$, i.e., $\mathbb{E}[W^{a}_m]<\infty$ for all $a\in(0,1)$. Note that the scheme in \eqref{Eqn:APAssoScheme} makes users associate with an AP in any tier no matter which spectrum the AP primarily/only uses, and it can cover several different pathloss-based AP association schemes by changing the design of the AP association weights, such as the \textit{biased nearest AP association} (BNA) scheme if $W_{m}$ is a constant, the \textit{biased mean strongest AP association} (BMSA) scheme if $W_{m_i}\equiv  b_mP_mH^{(s)}_{m_i}$ for all $m\in\mathcal{M}$ where $b_m>0$ is a constant bias and $H^{(s)}_{m_i}$  characterizes the large-scale channel gain of the tier-$m$ APs such as shadowing, and other schemes etc. \cite{CHLLCW1502}\cite{CHLLCW16}.

\subsection{AP Association Probability and Cell Load Statistics}
The AP association scheme in \eqref{Eqn:APAssoScheme} can be reformulated to statistically represent the weighted pathloss of AP $X_o$ given by 
\begin{align}\label{Eqn:WeightedPathLoss}
W_o\|X_o\|^{-\alpha}\stackrel{d}{=}\sup_{X_{m_i}\in\bigcup_{m=1}^M\X_m} W_{m_i}\|X_{m_i}\|^{-\alpha},
\end{align} 
where $W_o\in\{W_{m_i}, m\in\mathcal{M}, i\in\mathbb{N}_+\}$ is the AP association weight used by AP $X_o$ and $\stackrel{d}{=}$ means the statistical equivalence in distribution. The statistical property of $W_o\|X_o\|^{-\alpha}$ is crucial for the following analysis and it is provided in the following lemma.
\begin{lemma}[Revised from \cite{HSJYJSXPJGA12,CHLKLF16}]\label{Lem:AssDistPDF}
Let $\|\widetilde{X}_o\|^{-\alpha}$ denote the weighted pathloss of the AP $X_o$ given in \eqref{Eqn:WeightedPathLoss}, i.e., $\|\widetilde{X}_o\|^{-\alpha}=W_o\|X_o\|^{-\alpha}$. The probability density function (pdf) of $\|\widetilde{X}_o\|$ can be shown as
\begin{align}\label{Eqn:pdfDisAssoAP}
f_{\|\widetilde{X}_o\|}(x)= 2\pi x \widetilde{\lambda}\exp\left(-\pi\widetilde{\lambda} x^2\right),
\end{align}
where $\widetilde{\lambda}\defn  \sum_{m=1}^{M}\widetilde{\lambda}_m$ and $\widetilde{\lambda}_m\defn \lambda_m\mathbb{E}\left[W^{\frac{2}{\alpha}}_m\right]$.
\end{lemma}
\begin{IEEEproof}
The proof of \eqref{Eqn:pdfDisAssoAP} can be referred to \cite{HSJYJSXPJGA12,CHLKLF16}.
\end{IEEEproof}

According to the proof of Lemma \ref{Lem:AssDistPDF}, we essentially realize that $\|\widetilde{X}_o\|=W^{-1/\alpha}_o\|X_o\|$ can be statistically viewed as the distance from the nearest point in the PPP of $\bigcup_{m\in\mathcal{M}}\widetilde{X}_m$ to typical user $U_0$ \cite{DSWKJM96,CHLLCW16,CHLKLF16}. This fact implies the probability that users associate with a tier-$m$ AP is given by
\begin{align}\label{Eqn:AssProbabilityTierm}
\vartheta_m=\frac{\lambda_m\mathbb{E}\left[W^{\frac{2}{\alpha}}_m\right]}{\sum_{k=1}^{M}\lambda_k\mathbb{E}\left[W^{\frac{2}{\alpha}}_k\right]}=\frac{\widetilde{\lambda}_m}{\widetilde{\lambda}}=\frac{\lambda_m\bar{w}^{\frac{2}{\alpha}}_m}{\sum_{k=1}^{M}\lambda_k\bar{w}^{\frac{2}{\alpha}}_k},
\end{align}
which can be used to characterize the distribution of the number of users associating with an AP in a particular tier as shown in the following lemma. 
\begin{lemma}\label{Lem:CellLoadPDF}
Let $\mathcal{A}_{m_i}$ denote the cell area where all users associate with AP $X_{m_i}$ by using the AP association scheme \eqref{Eqn:APAssoScheme}. If the cell load of a tier-$m$ AP, denoted as $\X_m(\mathcal{A}_{m_i})$, is defined as the number of users associating a tier-$m$ AP and $\mathbb{E}\left[W^{2/\alpha}_m\right]\mathbb{E}\left[W^{-2/\alpha}_m\right]<\infty$, then its probability mass function (pmf) is given by
\begin{align}\label{Eqn:CellLoadPDF}
\nu_{m,n}\defn\mathbb{P}\left[\X_m(\mathcal{A}_{m})=n\right]=\frac{\Gamma(n+\zeta_m)}{n!\Gamma(\zeta_m)}\left(\frac{\zeta_m\lambda_m}{\zeta_m\lambda_m+\mu\vartheta_m}\right)^{n+\zeta_m},\, n\in\mathbb{N},
\end{align}
where $\Gamma(x)=\int_{0}^{\infty} t^{x-1}e^{-t}\dif t$ is the Gamma function, $\zeta_m\defn \frac{7}{2}\mathbb{E}\left[W^{2/\alpha}_m\right]\mathbb{E}\left[W^{-2/\alpha}_m\right]$ and $\widetilde{\lambda}$ is already defined in Lemma \ref{Lem:AssDistPDF}.
\end{lemma}
\begin{IEEEproof}
Since users adopt the scheme in \eqref{Eqn:APAssoScheme} to select their serving AP and $\widetilde{X}_o$ can be viewed as the nearest point in $\bigcup_{m=1}^M \widetilde{\X}_m$ to the typical user, cell load $\X_m(\mathcal{A}_{m_i})$ has the same distribution as $\widetilde{X}_m(\widetilde{\mathcal{A}}_{m_i})$ where $\widetilde{\mathcal{A}}_m$ is the cell area of AP $\widetilde{X}_{m_i}\in\widetilde{\X}_m$ and it is Voronoi-tessellated. Hence, it follows that
\begin{align*}
\nu_{m,n}=\mathbb{P}\left[\X_m(\mathcal{A}_{m})=n\right]=\mathbb{P}\left[\widetilde{\X}_m(\widetilde{\mathcal{A}}_{m})=n\right]=\mathbb{E}_{\widetilde{\mathcal{A}}_m}\left[ \frac{(\mu\widetilde{\lambda}_m\widetilde{\mathcal{A}}_m/\widetilde{\lambda})^n}{n!}e^{-\mu\widetilde{\lambda}_m\widetilde{\mathcal{A}}_m/\widetilde{\lambda}}\right].
\end{align*}
Since the exact pdf of a voronoi-tessellated area is still an open problem, its substitute expression accurately approximated by a Gamma random variable can be found as \cite{CHLLCW1502,CHLLCW16}
\begin{align*}
f_{\widetilde{\mathcal{A}}_m}(x) = \frac{(\zeta_m\lambda_mx)^{\zeta_m}}{x\Gamma(\zeta_m)}e^{-\zeta_m\lambda_mx}.
\end{align*} 
Thus,
\begin{align*}
\nu_{m,n}=\frac{1}{n!}\left(\frac{\mu\widetilde{\lambda}_m}{\widetilde{\lambda}}\right)^n\frac{(\zeta_m\lambda_m)^{\zeta_m}}{\Gamma(\zeta_m)}\int_{0}^{\infty} x^{n+\zeta_m-1} e^{-(\mu\widetilde{\lambda}_m/\widetilde{\lambda}+\zeta_m\lambda_m)x},
\end{align*}
and substituting the result in \eqref{Eqn:AssProbabilityTierm} into $\nu_{m,n}$ yields the result in \eqref{Eqn:CellLoadPDF}.
\end{IEEEproof}

The pmf of the cell load in Lemma \ref{Lem:CellLoadPDF} indicates that the probability of no users associating a tier-$m$ AP is
\begin{align}\label{Eqn:VoidProb}
\mathbb{P}[V_m=0]=\nu_{m,0}=\left(1+\frac{\mu\vartheta_m}{\zeta_m\lambda_m}\right)^{-\zeta_m},
\end{align}
which is the void probability of a tier-$m$ AP. As $\zeta_m$ (or $\mathbb{E}[W^{2/\alpha}_m]$) goes to infinity, it reduces to the following minimum:
\begin{align}
\lim_{\zeta_m\rightarrow\infty} \nu_{m,0} = \exp\left(-\frac{\mu}{\lambda_m}\right).
\end{align}
Thus, the void probabilities of the APs cannot be ignored if the ratios of the user intensity to the AP intensities are not large, especially for the network with densely deployed APs.  However, most prior works on the modeling of the multi-tier PPP-based HetNets overlook this important void AP  issue.  Also, the mean cell load of a tier-$m$ AP is given by
\begin{align}
\mathbb{E}\left[\X_m(\mathcal{A}_m)\right]=\frac{\widetilde{\lambda}_m}{\widetilde{\lambda}}\mu\times \frac{1}{\lambda_m}=\frac{\mu\mathbb{E}\left[W^{\frac{2}{\alpha}}_m\right]}{\sum_{k=1}^{M}\lambda_k\mathbb{E}\left[W^{\frac{2}{\alpha}}_k\right]}=\frac{\mu \bar{w}^{\frac{2}{\alpha}}_m}{\sum_{k=1}^{M}\lambda_k \bar{w}^{\frac{2}{\alpha}}_k}.\label{Eqn:CellLoadAPmthTier}
\end{align}
Hence, to completely balance the mean load between all tiers, all $W_m$'s must be i.i.d. for all $m\in\mathcal{M}$. For example, the (unbiased) nearest AP association scheme that makes user associate their nearest AP can achieve a completely balanced and same mean cell load for all APs in different tiers in that all $W_{m_i}$'s in the scheme are the same constant.

\subsection{Channel Access Protocols}
In this paper, all APs are assumed to always have data to transmit to their tagged users. The channel access protocols for the licensed spectrum and unlicensed spectrum are quite different. All APs in the first $M-1$ tiers share the entire licensed spectrum at the same time and they are synchronized when accessing the licensed channel\footnote{Such a licensed channel access protocol is widely used in the cellular networks. In addition, we assume there is only one channel in the licensed spectrum for the ease of analysis.}.  Note that the APs in the $M$th tier cannot access the licensed channel and they are only allowed to access the channel in the unlicensed spectrum. All APs have to use the (slotted non-persistent) \textit{opportunistic} CSMA/CA (carrier sense multiple access with collision avoidance) protocol to access the unlicensed channel\footnote{Like the case in the licensed spectrum, we assume there is only one available channel in the unlicensed spectrum  in order to simplify our following analysis.}. By adopting such an opportunistic CSMA/CA protocol, the APs whose channel gains are greater than some threshold are qualified and synchronized to contend the unlicensed channel in the predesignated time slots. The feature of this opportunistic CSMA/CA protocol is able to make the unlicensed spectrum resource to be utilized effectively by the APs with good channel conditions so as to improve the spectrum sharing efficiency and throughput. Each AP in the $m$-th tier that performs the opportunistic CSMA/CA protocol has a sensing region $\mathcal{S}_m$ in which all unlicensed channel accessing activities can be detected by the AP. The channel access probability of the opportunistic CSMA/CA protocol is already derived in our previous works \cite{CHLHCT16}\cite{HCTCHL16}, and it can be modified for the AP association scheme in \eqref{Eqn:APAssoScheme} and rewritten as shown in the following lemma.
\begin{lemma}[Channel access probability in the unlicensed spectrum modified from \cite{CHLHCT16}\cite{HCTCHL16}] \label{Lem:ChannelAccessProbability}
If the random backoff time of a tier-$m$ AP is uniformly distributed in $[0,\tau_m]$, then its channel access probability is
\begin{align}\label{Eqn:ChannelAssProb1}
\rho_m = \frac{1-e^{-\frac{\tau_M}{\mathbb{E}[W^{2/\alpha}_m]}\sum_{k=1}^{M}A_{m,k}\lambda^{\dag}_{k,M}}}{\tau_m\sum_{k=1}^{M}A_{m,k}(\bar{w}_k/\bar{w}_m)^{\frac{2}{\alpha}}\lambda^{\dag}_{k,M}}+\sum_{j=m}^{M} \frac{e^{-\tau_j\sum_{k=1}^{j}A_{m,k}(\frac{\bar{w}_k}{\bar{w}_m})^{\frac{2}{\alpha}}\lambda^{\dag}_{k,j}}-e^{-\tau_{j+1}\sum_{k=1}^{j}A_{m,k}(\frac{\bar{w}_k}{\bar{w}_m})^{\frac{2}{\alpha}}\lambda^{\dag}_{k,j}}}{\tau_m\sum_{k=1}^{j}A_{m,k}(\bar{w}_k/\bar{w}_m)^{\frac{2}{\alpha}}\lambda^{\dag}_{k,j}},
\end{align}
where $\tau_1\geq\tau_2\geq \cdots\geq \tau_M\geq 0$, $\tau_{M+1}\equiv 0$, $A_{m,k}$ is the mean area of region $\mathcal{S}_m$ where the tier-$k$ APs are distributed (see \cite{CHLHCT16} for the details of how to calculate $A_{m,k}$.), $\lambda^{\dag}_{k,j}=\xi_k(1-\nu_{k,0})\lambda_k(\frac{\tau_j-\tau_{j+1}}{\tau_k})$, $\xi_k\in[0,1]$ is the probability that the unlicensed channel (power) gain from a tier-$k$ AP to its servicing user is greater than threshold $\delta>0$. If all $\tau_m$'s are the same and equal to $\tau$, \eqref{Eqn:ChannelAssProb1} reduces to
\begin{align}
\rho_m=\frac{1-\exp\left(-\tau\sum_{k=1}^{M}A_{m,k}\xi_kq_{k,0}\lambda_k(\bar{w}_k/\bar{w}_m)^{\frac{2}{\alpha}}\right)}{\tau\sum_{k=1}^{M}A_{m,k}\xi_kq_{k,0}\lambda_k(\bar{w}_k/\bar{w}_m)^{\frac{2}{\alpha}}},
\end{align}
where $q_{k,0}\defn 1-\nu_{k,0}$ is the non-void probability of the tier-$k$ APs.
\end{lemma}

The channel access probability in \eqref{Eqn:ChannelAssProb1} indicates not only how much chance a tier-$m$ AP can successfully access the unlicensed channel in a particular timeslot but also the fraction of time it can access the unlicensed channel in the long-term sense. Adjusting the random backoff time limit $\tau_m$ can make the tier-$m$ APs have more/less priority or time fraction to access the unlicensed spectrum. For example, if the tier-$M$ APs represent the WiFi APs, we can make its backoff time limit $\tau_M$ much shorter than those of the APs in the first $M-1$ tiers so that the throughput of the WiFi APs is guaranteed to remain at some level and not significantly reduced when the unlicensed spectrum is shared by many APs in other tiers at the same time. This is similarly implementing the ideas of the Listen-Before-Talk (LBT) with Carrier Sensing Adaptive Transmission (CSAT) and Licensed-Assisted Access (LAA) protocols proposed in the LTE-U \cite{HZXCWGSW15,RZMWLXCZZXSLLX15}. Another two characteristics of the result in \eqref{Eqn:ChannelAssProb1} are including the  probability of being void APs  as well as the probability of opportunistically having a good channel state. They make the channel access probability more accurate and higher for the APs with good channels. In the following analysis, we will see how the channel access probability plays a pivotal role in analyzing the throughput in the unlicensed spectrum.

\section{Shannon Transform, Mean Spectrum Efficiency and Per-User Throughput}\label{Sec:MeanSpec}
In this section, we would like to study the mean spectrum efficiencies of a user in the unlicensed and licensed spectra. Finding the explicit expressions of these mean spectrum efficiencies is an important task since they provide the insights into how to manage traffic flow in order to achieve the throughput optimality. The prior approaches to deriving the mean spectrum efficiency in the literature are based on integrating the function of the coverage (success) probability under the Rayleigh fading channel model\cite{JGAFBRKG11,HSJYJSXPJGA12,PXCHLJGA13}. Hence, these approaches cannot characterize the mean spectrum efficiencies in a non-Rayleigh fading environment. In the following analysis, we will show how to derive the mean spectrum efficiency for any general channel gain and AP association weight models in a low-complexity expression. First, we need to introduce the Shannon transform of a nonnegative random variable and its integral identity since they are the key to obtaining the analytically tractable expression of the mean spectrum efficiency without making any specific modeling assumptions on the channel gains and AP association weights.

\subsection{The Shannon Transform and Its Integral Identity}
The Shannon transform of a nonnegative random variable $\Psi$ for any nonnegative $\eta\in\mathbb{R}_{++}$ is defined as
\begin{align}\label{Eqn:DefnShannonTransform}
	\mathcal{S}_{\Psi}(\eta) = \mathbb{E}\left[\ln(1+\eta \Psi)\right].
\end{align}
The Shannon transform in \eqref{Eqn:DefnShannonTransform} has an identity as shown in the following theorem. 
\begin{theorem}[The integral identity of the Shannon transform]\label{Thm:ShannonTransform}
	Consider a nonnegative random variable $\Psi$ and the Laplace transform of its reciprocal is always well-defined and exists, i.e., $\mathcal{L}_{\Psi^{-1}}(s)<\infty$ for $s\in\mathbb{R}_{++}$. Define $\mathcal{S}_{\Psi}(\eta)\defn\mathbb{E}[\ln(1+\eta\Psi)]$ as the Shannon transform of random variable $\Psi$ for a nonnegative constant $\eta\in\mathbb{R}_+$. If $\mathcal{S}_{\Psi}(\eta)$ exists for any $\eta\in\mathbb{R}_+$, then it has the following identity
	\begin{align}\label{Eqn:IdenShannonTrans}
		\mathcal{S}_{\Psi}(\eta) =\int_{0^+}^{\infty}\frac{(1-e^{-\eta s})}{s}\mathcal{L}_{\Psi^{-1}}(s) \dif s, 
	\end{align}
	which always holds. Furthermore, we can have
	\begin{align}\label{Eqn:IdenShannonTrans2}
	\mathbb{E}\left[\mathcal{S}_{\Psi}(\eta)\right] =\int_{0^+}^{\infty}\frac{[1-\mathcal{L}_{\eta}(s)]}{s}\mathcal{L}_{\Psi^{-1}}(s) \dif s
	\end{align}
	if $\eta$ is a nonnegative random variable and its Laplace transform exists.
\end{theorem}
\begin{IEEEproof}
	See Appendix \ref{App:ProofShannonTransform}.
\end{IEEEproof}
\noindent The identity of the Shannon transformation in Theorem \ref{Thm:ShannonTransform} provides a low-complexity means to find the mean spectrum efficiency of a link if $\eta\Psi$ is the SIR of the link and the Laplace transform of their reciprocal exists and can be explicitly found. As we will show in the following subsections, Theorem \ref{Thm:ShannonTransform} facilitates the derivations of the mean spectrum efficiencies of an AP in the unlicensed and licensed spectra for any general random channel gain and AP association weight models. 

\subsection{Mean Spectrum Efficiency of the APs in the Unlicensed Spectrum}
Consider the scenario that typical user $U_0$ associates with a tier-$m$ AP by scheme \eqref{Eqn:APAssoScheme} and is receiving data over the unlicensed channel. The mean spectrum efficiency (bps/Hz) the tier-$m$ AP offers to the user in the unlicensed spectrum is defined as
\begin{align}\label{Eqn:MeanRateUnlicensed}
R_{\Ut_m}\defn\mathbb{E}\left[\log_2\left(1+\frac{P_mH_m}{I_{\Ut_m}\|X_o\|^{\alpha}}T_m\Xi_m\right)\right],\,m\in\mathcal{M},
\end{align}
where $P_m$ is the transmit power used by AP $X_o\in\X_m$, $H_m$ is the random channel (power) gain with mean $\bar{h}_m$ from AP $X_o$ to typical user $U_0$, and $I_{\Ut_m}$ is the interference as shown in the following
\begin{align*}
	I_{\Ut_m} \defn \sum_{X_{m_i}\in\bigcup_{m=1}^M \X_m\setminus X_o} P_mH_{m_i}V_{m_i}T_{m_i}\Xi_{m_i}\|X_{m_i}\|^{-\alpha}
\end{align*} 
in which $H_{m_i}$ denotes the random channel gain with mean $\bar{h}_m$ from AP $X_{m_i}$ to typical user $U_0$, $\Xi_{m_i}\in\{0,1\}$ is a Bernoulli random variable that equals to one if AP $X_{m_i}$'s channel gain is greater than threshold $\delta>0$ and zero otherwise, $T_{m_i}\in\{0,1\}$ is also a Bernoulli random variable that equals to one if AP $X_{k_i}$ can access the unlicensed channel and zero otherwise\footnote{Namely, the probability that $\Xi_{m_i}$ is equal to one is $\mathbb{P}[\Xi_{m}=1]=\xi_m$, and the probability that $T_{m_i}$ is equal to one is $\mathbb{P}[T_{m}=1]=\rho_m$ for all $m\in\mathcal{M}$.}, $T_m\in\{T_{m_i}\}$ indicates whether AP $X_o$ can access the unlicensed channel, and $\Xi_m\in\{\Xi_{m_i}\}$ indicates whether  the channel gain of AP $X_o$ in the unlicensed spectrum is greater than threshold $\delta$. For the ease of analysis, we assume that all $\frac{H_{m_i}}{\bar{h}_m}$'s are i.i.d. for all $i\in\mathbb{N}_+$ and $m\in\mathcal{M}$ throughout this paper.

Letting $\Psi_{\Ut_m}\defn P_mH_mT_m\Xi_m/I_{\Ut_m}\|X_o\|^{\alpha}$ and using the definition of the Shannon transform in the previous subsection, $R_{\Ut_m}$ can be further expressed as
\begin{align}\label{Eqn:MeanRateUnlicensed2}
R_{\Ut_m}=\frac{\rho_m\xi_m}{\ln(2)}\int_{0^+}^{\infty} \frac{1-\mathcal{L}_{H_m}(s)}{s}\mathcal{L}_{P^{-1}_m\|X_o\|^{\alpha}I_{\Ut_m}}(s) \dif s
\end{align}
and the explicit expression of $R_{\Ut_m}$ derived by using Theorem \ref{Thm:ShannonTransform} is shown in the following theorem.
\begin{theorem}\label{Thm:LowBoundUnlicensedMeanRate}
Suppose users adopt the AP association scheme in \eqref{Eqn:APAssoScheme}. If all non-void APs use the opportunistic CSMA/CA protocol to access the unlicensed channel, the mean spectrum efficiency of the user in the unlicensed spectrum in \eqref{Eqn:MeanRateUnlicensed2} can be explicitly lower bounded by
\begin{align}
R_{\Ut_m} \geq \frac{\rho_m\xi_m}{\ln (2)} \bigintsss_{0^+}^{\infty} \frac{\left[1-\mathcal{L}_{\widehat{H}}(u)\right]\dif u}{u\left(\sum_{k=1}^{M}q_{k,0}\xi_k\rho_k\vartheta_k\ell_{\widehat{H}}(\frac{\bar{w}_m\bar{h}_kP_k}{\bar{w}_k\bar{h}_mP_m}u,\frac{2}{\alpha})+1\right)}, \label{Eqn:UnlicensedMeanRate}
\end{align}
where $\widehat{H}\defn \frac{H_m\bar{w}_m}{W_m\bar{h}_m}$ is a random variable with unit mean and $\ell_Z(x,y)$ for $y\in(0,1)$ is defined as
\begin{align}
\ell_{Z}(x,y)\defn x^y\Gamma\left(1-y\right)\mathbb{E}\left[Z^y\right]+\int_{0}^{1}\mathcal{L}_Z\left(x t^{-\frac{1}{y}}\right)\dif t-1.
\end{align}
If all $W_{m_i}$'s are deterministic and equal to constant $\bar{w}_m$ and all $H_{m_i}$'s are i.i.d. random variables with unit mean for all $m\in\mathcal{M}$, then \eqref{Eqn:UnlicensedMeanRate} reduces to
\begin{align}
R_{\Ut_m} \geq \frac{\rho_m\xi_m}{\ln (2)}\bigintsss_{0^+}^{\infty} \frac{\left[1-\mathcal{L}_{H}(u)\right]\dif u}{u\left(\sum_{k=1}^{M}q_{k,0}\xi_k\rho_k\vartheta_k\ell_{H}(\frac{\bar{w}_m\bar{h}_kP_k}{\bar{w}_k\bar{h}_mP_m}u,\frac{2}{\alpha})+1\right)}, \label{Eqn:UnlicensedMeanRate2}
\end{align}
where $H$ has the same distribution as all $H_{m_i}$'s.
\end{theorem}
\begin{IEEEproof}
See Appendix \ref{App:ProofLowBoundUnlicensedMeanRate}.
\end{IEEEproof}

It is worth mentioning a few features of the lower bounds in \eqref{Eqn:UnlicensedMeanRate} and \eqref{Eqn:UnlicensedMeanRate2}. First of all, the lower bound is in general \textit{fairly tight} since the location correlations between the non-void APs due to AP association and opportunistic CSMA/CA are usually very weak. These location correlations will be weakened and thus $R_{\Ut_m}$ will be very close to its lower bound when either the user intensity or the channel gain threshold for opportunistic CSMA/CA increases. The tightness of the lower bound in \eqref{Eqn:UnlicensedMeanRate} will be verified by the numerical results presented in Section \ref{Subsec:SimMeanSpecEfff}. Second, the lower bound in \eqref{Eqn:UnlicensedMeanRate} is valid for all random models of channel gains and AP association weights  as long as the Laplace transforms of the channel gains and the AP association weights exist, which is never derived in the literatures. This is a very important feature since we are able to realize how different channel and AP association models affect $R_{\Ut_m}$ and gain some insights about how to improve $R_{\Ut_m}$ by appropriately designing the AP association weights in order to manage the traffic flows among different tiers of the APs. Third, according to Jensen's inequality, we know $\mathcal{L}_Z(u)\geq \exp(-u\mathbb{E}[Z])$, which means if all $H_m$'s and/or $W_m$'s are not random the lower bound in \eqref{Eqn:UnlicensedMeanRate} will increase (i.e., $R_{\Ut_m}$ will increase). For example, in the special case that there are no random channel gain impairments and users adopt the BNA scheme (i.e., $H_m\equiv 1$ and all $W_{m_i}$'s are equal to constant $w_m$ for all $i\in\mathbb{N}_+$) or in the special case of the AP association scheme with $W_{m_i}=b_mP_mH_{m_i}$ for all $m\in\mathcal{M}$, the lower bound in \eqref{Eqn:UnlicensedMeanRate2} becomes
\begin{align}
R_{\Ut_m} \geq \frac{\rho_m\xi_m}{\ln (2)}  \bigintsss_{0^+}^{\infty} \frac{\left(1-e^{-u}\right)\dif u}{u\left(\ell(u,\frac{2}{\alpha})\sum_{k=1}^{M}q_{k,0}\xi_k\rho_k\vartheta_k+1\right)}, \label{Eqn:UnlicensedMeanRate3}
\end{align}
where $\ell(x,y)\defn x^y\Gamma\left(1-y\right)+\int_{0}^{1}e^{-x t^{-\frac{1}{y}}}\dif t-1$. And this is the maximum lower bound achieved by the biased AP association schemes in that the randomness of the channel gains does not exist in the $R_{\Ut_m}$.

\subsection{Mean Spectrum Efficiency of the APs in the Licensed Spectrum}
For the mean spectrum efficiency of a user associating with a tier-$m$ AP in the licensed spectrum, its formal definition can be written as
\begin{align}\label{Eqn:DefnMeanRateLicensed}
R_{\Lt_m}\defn\mathbb{E}\left[\log_2\left(1+\frac{P_mH_m}{I_{\Lt_m}\|X_o\|^{\alpha}}\right)\right],
\end{align}
where $X_o\in\X_m$ and $I_{\Lt_m}$ is given by
\begin{align}
I_{\Lt_m} \defn \sum_{X_{m_i}\in\bigcup_{m=1}^{M-1} \X_m\setminus X_o} P_mH_{m_i}V_{m_i}\|X_{m_i}\|^{-\alpha}.
\end{align} 
Note that all channel gains in \eqref{Eqn:DefnMeanRateLicensed} are evaluated in the licensed spectrum. The explicit result of $R_{\Lt}$ is shown in the following theorem. 
\begin{theorem}\label{Thm:LowBoundMeanRateLicensedBand}
If users adopt the AP association scheme in \eqref{Eqn:APAssoScheme}, then the mean spectrum efficiency of a user in the licensed spectrum defined in \eqref{Eqn:DefnMeanRateLicensed} can be shown as
\begin{align}
R_{\Lt_m}\geq  \frac{1}{\ln (2)} \bigintsss_{0^+}^{\infty} \frac{\left[1-\mathcal{L}_{\widehat{H}}(u)\right]\dif u}{u\left(\sum_{k=1}^{M-1}q_{k,0}\vartheta_k\ell_{\widehat{H}}(\frac{\bar{w}_m\bar{h}_kP_k}{\bar{w}_k\bar{h}_mP_m}u,\frac{2}{\alpha})+1\right)},\,m\in\{1,2,\ldots,M-1\}. \label{Eqn:MeanRateLincesed2}
\end{align}
If all $W_{m_i}$'s are deterministic and equal to $\bar{w}_m$ and all $H_{m_i}$'s are i.i.d. random variables with unit mean for all $m\in\mathcal{M}$, \eqref{Eqn:MeanRateLincesed2} reduces to
\begin{align}
R_{\Lt_m}\geq  \frac{1}{\ln (2)}  \bigintsss_{0^+}^{\infty} \frac{\left[1-\mathcal{L}_{H}(u)\right]\dif u}{u\left(\sum_{k=1}^{M-1}q_{k,0}\vartheta_k\ell_{H}(\frac{\bar{w}_m\bar{h}_kP_k}{\bar{w}_k\bar{h}_mP_m}u,\frac{2}{\alpha})+1\right)}. \label{Eqn:MeanRateLincesed3}
\end{align}
\end{theorem}
\begin{IEEEproof}
First, consider the case that all $W_{m_i}$'s are random variables. Since all APs in the first $M-1$ tiers access the licensed spectrum without using the opportunistic CSMA/CA protocol, the mean spectrum efficiency for a user associating with a tier-$m$ AP can be explicitly obtained as
\begin{align}
\mathbb{E}\left[\log_2\left(1+\frac{P_mH_m}{I_{\Lt_m}\|X_o\|^{\alpha}}\right)\right] =\frac{1}{\ln(2)}\bigintsss_{0^+}^{\infty}\frac{\left[1-\mathcal{L}_{\widehat{H}_m}(u)\right]\dif u}{u\left(\sum_{k=1}^{M-1}q_{k,0}\vartheta_k\ell_{\widehat{H}_k}(\frac{P_k\bar{w}_m\bar{h}_k}{P_m\bar{w}_k\bar{h}_m}u,\frac{2}{\alpha})+1\right)}\label{Eqn:mTierMeanRate}
\end{align}
by applying the result of $R_{\Ut_m}$ for the mean spectrum efficiency in the $m$-th tier with $\rho_k=\xi_k=1$ for all $k\in\mathcal{M}$. Substituting \eqref{Eqn:mTierMeanRate} into \eqref{Eqn:DefnMeanRateLicensed} results in \eqref{Eqn:MeanRateLincesed2}.
\end{IEEEproof}

The results in Theorem \ref{Thm:LowBoundMeanRateLicensedBand} are valid for any channel gain and AP association weight models and they are  never derived in the literature as well. Prior results on the spectrum efficiency in the licensed spectrum are only derived for Rayleigh fading channels and do not reflect the impact of the void APs. The lower bound on $R_{\Lt_m}$ in \eqref{Eqn:MeanRateLincesed2} is obtained by assuming the non-void correlated APs form $M$ independent thinning homogeneous PPPs and in general it is also very tight, like the lower bounds on $R_{\Ut_m}$ in Theorem \ref{Thm:LowBoundUnlicensedMeanRate}, since the location correlations of the non-void APs are fairly weak. Thus, as the user intensity goes to infinity, $q_{k,0}$ converges to one so that $R_{\Lt_m}$ exactly reduces to the lower bound given by
\begin{align}
\lim_{\mu\rightarrow\infty} R_{\Lt_m}=\frac{1}{\ln (2)} \bigintsss_{0^+}^{\infty} \frac{\left[1-\mathcal{L}_{\widehat{H}_m}(u)\right]\dif u}{u\left(\sum_{k=1}^{M-1}\vartheta_k\ell_{\widehat{H}_k}(\frac{P_k\bar{w}_m\bar{h}_k}{P_m\bar{w}_k\bar{h}_m}u,\frac{2}{\alpha})+1\right)},
\end{align}
which is the lowest mean spectrum efficiency of the tier-$m$ APs in the licensed spectrum. This indicates that \textit{the mean spectrum efficiency is significantly underestimated in a dense network if the void AP impact is not considered in the interference model}. Also, if the AP association scheme in \eqref{Eqn:APAssoScheme} with $W_{m_i}=b_mP_mH_{m_i}$ can characterize the channel gains, the lower bound in \eqref{Eqn:MeanRateLincesed2} becomes
\begin{align}
R_{\Lt_m}\geq  \frac{1}{\ln (2)} \bigintsss_{0^+}^{\infty} \frac{\left(1-e^{-u}\right)\dif u}{u\left(\ell(u,\frac{2}{\alpha})\sum_{k=1}^{M-1}q_{k,0}\vartheta_k+1\right)}, \label{Eqn:MeanRateLincesed4}
\end{align}
which is the maximum lower bound achieved by AP association and also the lower bound for the scenario that channels do not suffer random impairments and users select their APs by using the BNA scheme.  

\subsection{Per-User Throughput Characterization}  
We have characterized the mean spectrum efficiencies of the licensed and unlicensed spectra in the previous subsections, which can be used to characterize the total (licensed and unlicensed) link throughput of a tier-$m$ AP. Assume the bandwidths of the licensed spectrum and unlicensed spectrum are denoted by $B_{\Lt}$ and $B_{\Ut}$, respectively. Accordingly, the total link throughput of a tier-$m$ AP can be expressed as
\begin{align}\label{Eqn:ErgodicLinkRate}
C_m=B_{\Lt}R_{\Lt_m}\mathds{1}(m\neq M)+B_{\Ut}R_{\Ut_m},\, m\in\mathcal{M},
\end{align} 
where $\mathds{1}(\mathcal{E})$ is an indicator function which is one if event $\mathcal{E}$ is true and zero otherwise. Note that the link throughput of a tier-$M$ AP,  only in the unlicensed spectrum, is $C_M=B_{\Ut}R_{\Ut_M}$. The total link throughput of each AP highly depends on how the AP association weights in \eqref{Eqn:APAssoScheme} are designated. For instance, when $\mathbb{E}[W^{2/\alpha}_m]$ becomes larger, more users associate the tier-$m$ APs ( i.e., more traffic is offloaded to the tier-$m$ APs) so that the mean spectrum efficiencies would change very likely due to interference variations since the void probability of the tier-$m$ APs reduces and correspondingly the void probabilities of the APs in other tiers increase. In this case, whether the total link throughput of the tier-$m$ APs increases (or decreases) is dependent upon whether the interferences in the licensed and unlicensed spectra decrease (or increase). The channel access probability of the APs in the unlicensed spectrum decreases (or increases) due to offloading traffic to the tier-$m$ APs. 

Recall that the mean cell load of a tier-$m$ AP given in \eqref{Eqn:CellLoadAPmthTier}, denoted by $\mathbb{E}[\X_m(\mathcal{A}_m)]$, is the mean number of users associating a tier-$m$ AP. By assuming all users equally share the spectrum resources, the per-user link throughput of the tier-$m$ APs is given by
\begin{align}\label{Eqn:PerUserThroughputTierm}
c_m\defn\frac{C_m}{\mathbb{E}[\X_m(\mathcal{A}_m)|V_m=1]}=\frac{q_{m,0}\lambda_m}{\mu\vartheta_m}[B_{\Lt}R_{\Lt_m}\mathds{1}(m\neq M)+B_{\Ut}R_{\Ut_m}].
\end{align}  
In general, this per-user link throughput decreases as the traffic offloaded to the tier-$m$ APs increases in that $\mathbb{E}[\X_m(\mathcal{A}_m)]$ increases but $C_m$ may not increase.  Furthermore, we can define the per-user network throughput as $\sum_{m=1}^{M} c_m\vartheta_m$ since it can be explicitly written as
\begin{align}\label{Eqn:MeanPerUserLinkRate}
\sum_{m=1}^{M} c_m\vartheta_m=\frac{B_{\Lt}}{\mu}\sum_{m=1}^{M}q_{m,0}\lambda_m\left[R_{\Lt_m}\mathds{1}(m\neq M)+\frac{B_{\Ut}}{B_{\Lt}}R_{\Ut_m}\right]=\frac{1}{\mu}\sum_{m=1}^{M} q_{m,0}\lambda_mC_m,
\end{align}
and the term $\frac{1}{\mu}\sum_{m=1}^{M} q_{m,0}\lambda_mC_m$ characterizes how much network throughput a user can obtain when the void cell issue is considered and the unlicensed spectrum is opportunistically shared by all APs. In the following section, we will study how to maximize this per-user network throughput by optimizing the AP association weights in \eqref{Eqn:APAssoScheme}. 

\section{Traffic Management for Coexisting Licensed and Unlicensed APs}
The underlying HetNet considered in this paper models the scenario that licensed and unlicensed APs coexist and opportunistically share the unlicensed spectrum, which corresponds a realistic situation in which WiFi APs in the $M$th tier share their unlicensed spectrum with licensed LTE APs in the first $M-1$ tiers. The traffic management problem that interests us here is how to offload or load traffic between two orthogonal spectrum domains such that the per-user-based throughput increases. Since the APs in the first $M-1$ tiers could simultaneously assess the unlicensed channel, their transmitting behaviors in the unlicensed spectrum definitely affect the throughput of the APs in the $M$th tier. Accordingly, the primary premise of managing the traffic in different tiers is to make the per-user link throughput of the APs in the $M$th tier higher than some minimum required value. Namely, assuming the per-user link throughput of the tier-$M$ APs must be at least greater than some minimum value $c_{\min}$, i.e., $c_M\geq c_{\min}$ and it is 
\begin{align}
\frac{q_{M,0}B_{\Ut}}{c_{\min}\mu}\left(\sum_{m=1}^{M}\lambda_m\bar{w}^{\frac{2}{\alpha}}_m\right) \geq\frac{\bar{w}^{\frac{2}{\alpha}}_M}{R_{\Ut_M}}.\label{Eqn:TierMRateConstraint}
\end{align}
With this constraint on $\bar{w}^{\frac{2}{\alpha}}_M$, we are able to study how to maximize the per-user link throughput of the APs in any particular tier and the per-user network throughput by optimizing the designs of the AP association weights. 





\subsection{Decentralized Traffic Management}
According to Theorems \ref{Thm:LowBoundUnlicensedMeanRate},  \ref{Thm:LowBoundMeanRateLicensedBand} and the per-user link throughput of the tier-$m$ APs given in \eqref{Eqn:PerUserThroughputTierm}, $c_m$ is significantly affected by all $\{\bar{w}^{2/\alpha}_m\}$. Now our interest here is to gain some insights on when an AP in a particular tier should independently determine to offload its traffic to APs in other tiers (i.e., reduce its AP association weight) or load traffic from APs in other tiers (i.e., increase its AP association weight) so that its per-user link throughput increases. This is essentially a ``decentralized" traffic management problem since the traffic loading or offloading decision is independently made by each AP from the perspective of the per-user link throughput. In other words, this decentralized traffic management problem is to study how to increase or even maximize $c_m$ by unilaterally changing or even optimizing the value $\bar{w}^{2/\alpha}_m$ of a tier-$m$ AP under the constraint \eqref{Eqn:TierMRateConstraint}. That is, if possible, we would like to solve the following optimization problem of $\bar{w}^{2/\alpha}_m$:
\begin{align}
\begin{cases}
\max\limits_{\omega_m>0} &c_m(\omega_m) \\
\text{s.t. }&\left(\sum_{k=1}^{M}\lambda_k\omega_k\right) \frac{R_{\Ut_M}}{\omega_M} \geq \frac{c_{\min}\mu}{q_{M,0}B_{\Ut}}
\end{cases}
\end{align}
in which we define $\omega_k\defn \bar{w}^{2/\alpha}_k$ to simplify the notation in the problem.  The solution of this optimization problem exists as shown in the following lemma.
\begin{lemma}\label{Lem:OptimalWeightTierm}
Let $\Omega_m$ be the feasible set of $\omega_m$ with the constraint \eqref{Eqn:TierMRateConstraint}, i.e., it is
\begin{align}
\Omega_m \defn \left\{\omega_m\in\mathbb{R}_{++}: \left(\sum_{k=1}^{M}\lambda_k\omega_k\right) \frac{R_{\Ut_M}}{\omega_M} \geq \frac{c_{\min}\mu}{q_{M,0}B_{\Ut}}, \omega_k>0, k\in\mathcal{M}\setminus m\right\}.
\end{align}
If $\Omega_m$ is nonempty, there exists a maximizer $\omega^*_m$ of $c_m$ over $\Omega_m$ for all $m\in\{1,2,\ldots, M-1\}$.
\end{lemma}
\begin{IEEEproof}
First, we  would like to show that $\Omega_m$ is a compact set if it is nonempty. According to \eqref{Eqn:ChannelAssProb1} and \eqref{Eqn:UnlicensedMeanRate}, we can infer the following facts
\begin{align*}
\lim_{\omega_m\rightarrow 0}\left(\sum_{k=1}^{M}\lambda_k\omega_k\right) \frac{R_{\Ut_M}}{\omega_M}<\infty \text{ and } \lim_{\omega_m\rightarrow \infty}\left(\sum_{k=1}^{M}\lambda_k\omega_k\right) \frac{R_{\Ut_M}}{\omega_M}=0.
\end{align*}
Since $\left(\sum_{k=1}^{M}\lambda_k\omega_k\right) \frac{R_{\Ut_M}}{\omega_M}$ is bounded, there must exist a positive number $R^{\dagger}_{\Ut_M}$ at $\omega^{\dagger}_m$ such that 
$$R^{\dagger}_{\Ut_M}(\omega^{\dagger}_m)=\sup_{\omega_m>0}\left\{\left(\sum_{k\in\mathcal{M}}\lambda_k\omega_k\right) \frac{R_{\Ut_M}}{\omega_M}\right\}\geq \frac{c_{\min}\mu}{q_{M,0}B_{\Ut}}$$ 
if $\Omega_m$  is nonempty. For $\omega_m\in(\omega^{\dagger}_m,\infty)$, $\left(\sum_{k=1}^{M}\lambda_k\omega_k\right) \frac{R_{\Ut_M}}{\omega_M}$ is a monotonically decreasing function of $\omega_m$. Accordingly, there must exist a $\omega^{\ddagger}_m$ such that $\left(\sum_{k=1}^{M}\lambda_k\omega_k\right) \frac{R_{\Ut_M}}{\omega_M}\geq \frac{c_{\min}\mu}{q_{M,0}B_{\Ut}}$ for $\omega_m\in (0,\omega^{\ddagger}_m]$. Hence $\Omega_m=(0,\omega^{\ddagger}_m]$ is closed and bounded. Furthermore, $c_m$ is a continuous function of $\omega_m$ over $\Omega_m$. According to the Weierstrass theorem \cite{DPB99}, the maximizer $\omega^*_m$ must exist. 
\end{IEEEproof}
\noindent Lemma \ref{Lem:OptimalWeightTierm} reveals that $\omega^*_m\in \{\arg\sup_{\omega_m\in\Omega_m} c_m(\omega_m)\}$ and $\frac{\partial c_m}{\partial \omega_m}|_{\omega_m=\omega^*_m}=0$. However, finding $\omega^*_m$ needs the information of other $\omega_k$'s which is in general unknown for the APs in the $m$-th tier in the decentralized context.

The two fundamental traffic management rules for a tier-$m$ AP can be easily realized as
\begin{align}\label{Eqn:BasicRuleTrafficManage}
\frac{\partial c_m}{\partial \omega_m}<0 \Leftrightarrow\begin{cases}
\text{loading traffic reduces $c_m$}\\
\text{offloading traffic increases $c_m$}
\end{cases}
\end{align} 
and
\begin{align}\label{Eqn:BasicRuleTrafficManage2}
\frac{\partial c_m}{\partial \omega_m}>0 \Leftrightarrow\begin{cases}
\text{loading traffic increases $c_m$}\\
\text{offloading traffic reduces $c_m$}
\end{cases}.
\end{align} 
These two rules indicate that \textit{the APs in the first $M-1$ tiers need to offload traffic if $\frac{\partial c_m}{\partial \omega_m}<0 $ and load traffic if $\frac{\partial c_m}{\partial \omega_m}>0 $ under the constraint that the APs in the $M$th tier need to maintain their per-user link throughput above  the threshold value $c_{\min}$}.  According to the facts in \eqref{Eqn:BasicRuleTrafficManage} and \eqref{Eqn:BasicRuleTrafficManage2}, we develop a decentralized traffic management scheme for the APs in each tier as shown in the following theorem. 
\begin{theorem}\label{Thm:DecenTrafficMang}
For the APs in the $m$-th tier and $m\in\{1,2,\ldots,M-1\}$, the following decentralized traffic management scheme can maximize their per-user link throughput under the constraint in \eqref{Eqn:TierMRateConstraint}
\begin{align}\label{Eqn:RecursiveWeightUpdate}
\omega_m(n+1) =\frac{ c^*(n)N_m(n)\omega_m(n)}{C_m(n)} ,\,\omega_m(0)>0,\, n\in\mathbb{N},
\end{align}
where $N_m(n)=\frac{1}{n}\sum_{i=0}^{n-1}N_m(i)$ denotes the average number of the users associating with a tier-$m$ AP at time $n$ and $c^*(n)\defn \max\{c_{\min},c^*(n-1),C_m(n)/N_m(n)\}$. In addition, as $n$ goes to infinity this scheme makes $\omega_m$ converge to $\omega^*_m$ that is the fixed point of the function $\Upsilon_m(x)$ given by
\begin{align}\label{Eqn:FunOptWeight}
\Upsilon_m(x)=\frac{q_{m,0}(x)C_m(x)}{\mu c_{\min}}\left(\sum_{k\in\mathcal{M}\setminus m}\lambda_k\omega_k+\lambda_mx\right).
\end{align}
\end{theorem}
\begin{IEEEproof}
See Appendix \ref{App:ProofDecenTrafficMang}.
\end{IEEEproof}

\noindent Since a tier-$m$ AP can estimate $R_{\Lt_m}(n)$ and $R_{\Ut_m}(n)$ and other parameters in \eqref{Eqn:RecursiveWeightUpdate} are locally available to the tier-$m$ AP, the scheme in \eqref{Eqn:RecursiveWeightUpdate} can be easily implemented by the AP. Function $\Upsilon_m(x)$ in \eqref{Eqn:FunOptWeight} can help us roughly determine the initial value $\omega_m(0)$ of $\omega_m(n)$ provided that each tier-$m$ AP initially knows all other $\omega_k$'s and this would shorten the  process of $\omega_m(n)$ converging to $\omega^*_m$. Note that in general the per-user link throughput achieved by the scheme in \eqref{Eqn:RecursiveWeightUpdate} in the steady state is just a suboptimal result because other $M-1$ parameters  $\{\omega_k, k\in\mathcal{M}\setminus m\}$ are not optimized. In the following subsection, a centralized traffic management approach to maximizing the per-user network throughput is proposed and studied.   

\subsection{Centralized Traffic Management}
If all the APs in the HetNet can send their information to their backhaul processing unit in their core network, the centralized traffic management is implementable\footnote{For example, in a cloud-RAN architecture the core network is able to know the information of all BSs so that it can perform the following proposed centralized traffic management scheme to maximize the per-user network throughput.}. Under this circumstance, we can maximize the per-user network throughput defined in \eqref{Eqn:MeanPerUserLinkRate} under the constraint in \eqref{Eqn:TierMRateConstraint} by optimizing all $\{\omega_m\}$. That is, we can solve the following optimization problem of the centralized traffic management 
\begin{align}\label{Eqn:AssWeightOptimization}
\begin{cases}
\max\limits_{\{\omega_m\}>0}&\sum_{m=1}^{M}q_{m,0}\lambda_m\left[R_{\Lt_m}\mathds{1}(m\neq M)+\frac{B_{\Ut}}{B_{\Lt}}R_{\Ut_m}\right]\\
\text{s.t. } &\left(\sum_{m=1}^{M}\lambda_m\omega_m\right) \frac{R_{\Ut_M}}{\omega_M} \geq \frac{c_{\min}\mu}{q_{M,0}B_{\Ut}} \\
&\omega_m> 0 
\end{cases}
\end{align} 
to find the optimal $M$-tuple vector of $(\omega_1,\omega_2,\ldots,\omega_M)$. Note that the objective function of the optimization problem in \eqref{Eqn:AssWeightOptimization} is the per-user network throughput normalized by constant $\frac{\mu}{B_{\Lt}}$ which does not affect the optimization solutions. The solution of this optimization problem exists as shown in the following theorem. 
\begin{lemma}
Let $\Omega$ be the feasible set of the $M$-tuple vector $(\omega_1,\omega_2,\ldots,\omega_M)$ with the constraint \eqref{Eqn:TierMRateConstraint} and it is expressed as
\begin{align}
\Omega\defn\left\{(\omega_1,\omega_2,\ldots,\omega_M)\in\mathbb{R}^M_{++}:\left(\sum_{m=1}^{M}\lambda_m\omega_m\right) \frac{R_{\Ut_M}}{\omega_M} \geq \frac{c_{\min}\mu}{q_{M,0}B_{\Ut}}\right\}.
\end{align}
If $\Omega$ is nonempty, there exists an optimal $M$-tuple vector $(\omega^*_1,\omega^*_2,\ldots,\omega^*_M)$ that maximizes the per-user network throughput  $\sum_{m=1}^{M} c_m\vartheta_m$.
\end{lemma}
\begin{IEEEproof}
According to Lemma \ref{Lem:OptimalWeightTierm}, we know there exists an $M$-tuple vector $(\omega_1,\ldots,\omega^*_m,\ldots,\omega_M)$ in set $\Omega_m$ if $\Omega_m$ is nonempty. In other words, there must exist a vector $(\omega^*_1,\omega^*_2,\ldots,\omega^*_M)\in \Omega_1\times\Omega_2\times\cdots\times\Omega_M\defn \prod_{m=1}^{M}\Omega_m$ if $\prod_{m=1}^{M}\Omega_m$ is nonempty since all $\Omega_m$'s are compact. Let $\Omega$ be an $M$-dimensional closed ball that encloses $\prod_{m=1}^{M}\Omega_m$, i.e., $\prod_{m=1}^{M}\Omega_m\subseteq\Omega$. Hence, $\Omega$ is compact as well as nonempty if $\prod_{m=1}^{M}\Omega_m$ is nonempty. Also, $\sum_{m=1}^{M} c_m\vartheta_m$ is continuous over $\prod_{m=1}^{M}\Omega_m$ in that $c_m$ is continuous over $\Omega_m$  for all $m\in\mathcal{M}$ and $\sum_{m=1}^{M} c_m\vartheta_m$ is a linear combination of all $c_m$'s, which follows that $\sum_{m=1}^{M} c_m\vartheta_m$ is also continuous over $\Omega$. Since $\sum_{m=1}^{M} c_m\vartheta_m$ is continuous over $\Omega$ and $\Omega$ is compact, there must exist an optimal $M$-tuple vector $(\omega^*_1,\omega^*_2,\ldots,\omega^*_M)$ that maximizes the per-user network throughput  $\sum_{m=1}^{M} c_m\vartheta_m$ according to the Weierstrass theorem.
\end{IEEEproof}
Although the optimal solution vector for the optimization problem in \eqref{Eqn:AssWeightOptimization} cannot be found in closed-form, we can resort to a numerical technique to acquire it. In addition, since the per-user network throughput can also be interpreted as the ``mean'' per-user link throughput of the HetNet as mentioned before, the solution of the optimization in \eqref{Eqn:AssWeightOptimization} essentially also maximizes the mean per-user link throughput of the HetNet.


\section{Numerical Example for Coexisting LTE and WiFi Networks}
\begin{table}[!t]
	\centering
	\caption{Network Parameters for Simulation}\label{Tab:SimPara}
	\begin{tabular}{|c|c|c|c|c|}
	\hline Parameter $\setminus$ AP Type (Tier \#)& Macrocell (1) & Picocell (2) & Femtocell (3) & WiFi (4)\\ \hline
	Power $P_m$ (W) & 20 & 1  & 0.2 & 0.1 \\ \hline
	Intensity $\lambda_m$ (APs/m$^2$) & $5\times 10^{-6}$ & $5\times 10^{-5}$ & $2.5\times 10^{-4}$ & $5\times 10^{-4}$   \\ \hline
	Maximum Backoff Time $\tau_m$ & $\infty$ &\multicolumn{2}{c|}{2}& 1 \\ \hline
	Sensing Area $\mathcal{S}_m$ (m$^2$) & N/A &\multicolumn{3}{c|}{900$\pi$}  \\ \hline
	CSMA Threshold $\delta$ & N/A & \multicolumn{3}{c|}{$4.481$}  \\ \hline  
	$H_{m_i}=H^{(f)}_m\times H^{(s)}_m$ & \multicolumn{4}{c|}{$\sim \exp(1,1)\times \ln\mathcal{N}(0,3(\text{dB}))$} \\ \hline
	Unlicensed Bandwidth $B_{\Ut}$ &\multicolumn{4}{c|}{160 MHz}\\ \hline 
	Licensed Bandwidth $B_{\Lt}$ &\multicolumn{4}{c|}{100 MHz}\\ \hline
	Pathloss Exponent $\alpha$ &\multicolumn{4}{c|}{4}\\ \hline 
	User Intensity $\mu$ (users/Km$^2$) &\multicolumn{4}{c|}{500}\\ \hline
	\end{tabular} 
\end{table}

In this section, we provide some numerical results by simulating a scenario that there are four tiers in the HetNet consisting of LTE  BSs and WiFi APs. The first tier of the HetNet consists of the macrocellcell BSs that do not access the unlicensed spectrum, the second and third tiers consist of picocell and femtocell BSs, and the fourth tier consists of the WiFi APs. The BSs and APs in the last three tiers use the opportunistic CSMA/CA protocol to access the unlicensed spectrum. All channels suffer Rayleigh fading and log-normal shadowing and all users adopt the BMSA scheme defined in Section \ref{Sec:SystemModel} to associate their serving BSs or APs. Specifically, the AP association weight in the scheme \eqref{Eqn:APAssoScheme} for the BSs in the first three tiers is designated as $W_{m_i}=bP_mH_{m_i}^{(s)}$ for $m\in\{1,2,3\}$ in which $b>0$ is a constant bias and $H_{m_i}^{(s)}$ characterizes the channel gain due to log-normal shadowing\footnote{In this section, the channel gain $H_{m_i}$ is equal to $H^{(f)}_{m_i}\times H^{(s)}_{m_i}$ where $H^{(f)}_{m_i}$ characterizes the channel gain due to Rayleigh fading and $H^{(s)}_{m_i}$, as already specified, characterizes the channel gain due to log-normal shadowing.}, whereas the AP association weight for the WiFi APs in the fourth tier is $W_{4_i}=H^{(s)}_{4_i}$, i.e., adopting the unbiased MSA scheme. Note that  we have $\omega_m=(bP_m\mathbb{E}[H^{(s)}_m])^{\frac{2}{\alpha}}$ for $m\in\{1,2,3\}$ and $\omega_4=(P_4\mathbb{E}[H^{(s)}_4])^{\frac{2}{\alpha}}$. All the network parameters for simulation are listed in Table \ref{Tab:SimPara}. In the following, we first provide the simulation results of the void BS/AP probability and the channel access probability for the proposed opportunistic CSMA/CA protocol. Next, the simulation results of the mean spectrum efficiencies in the licensed and unlicensed spectra are given to demonstrate the tightness of the derived lower bounds on the mean spectrum efficiencies in Section \ref{Sec:MeanSpec}. Finally, we show the simulation results of traffic offloading from the LTE network to the WiFi network in order to validate the previous discussions on the per-user link throughput of an AP and per-user network throughput with the decentralized and centralized traffic management schemes. 

\subsection{Simulation Results of the Void BS/AP Probability and the Channel Access Probability} 
\begin{figure}
\centering
\includegraphics[scale=0.35]{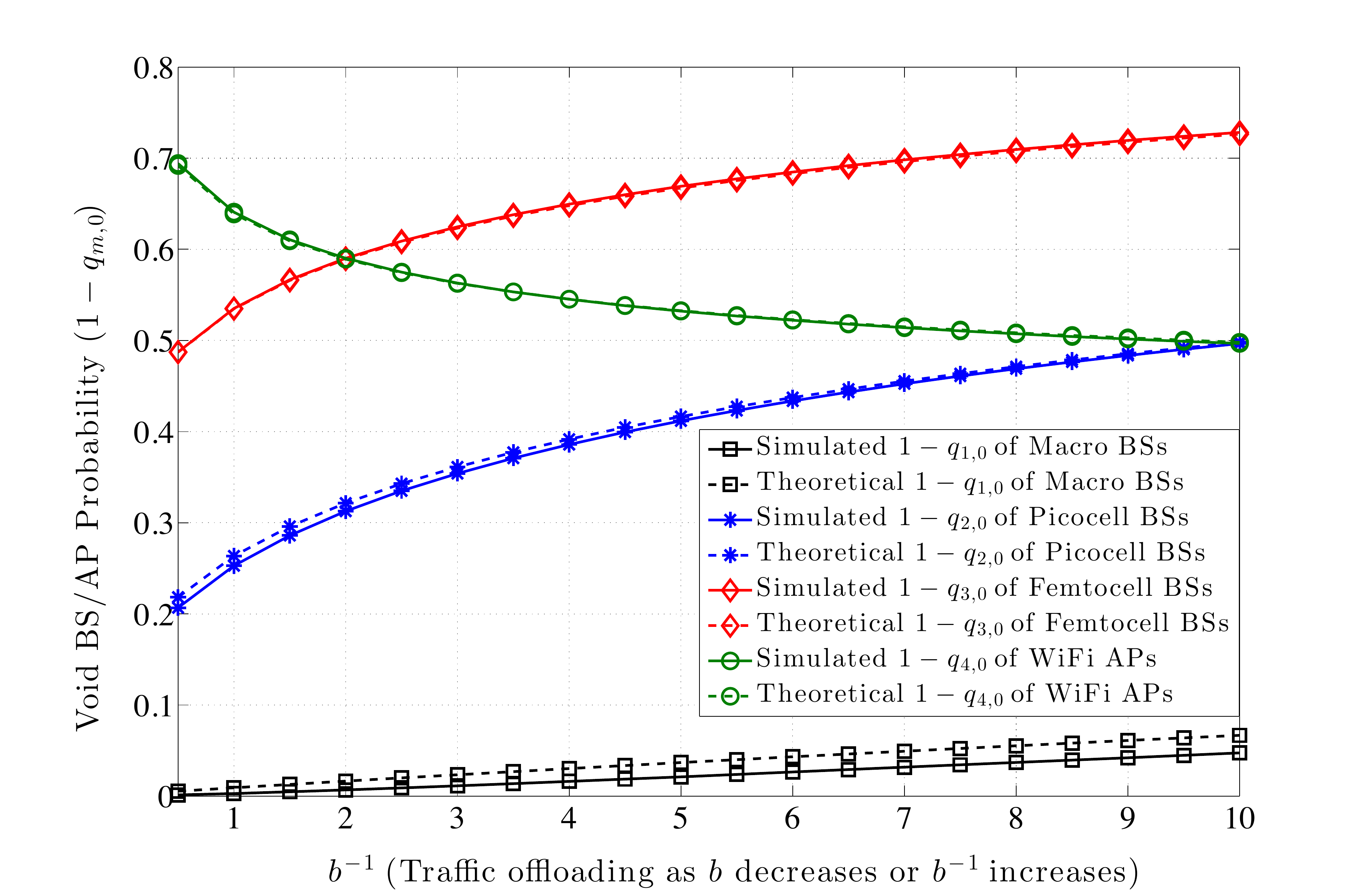}
\caption{Simulation results of the void probabilities of the BSs and APs in the four different tiers.}
\label{Fig:VoidProb}
\end{figure}

\begin{figure}
	\centering
	\includegraphics[scale=0.38]{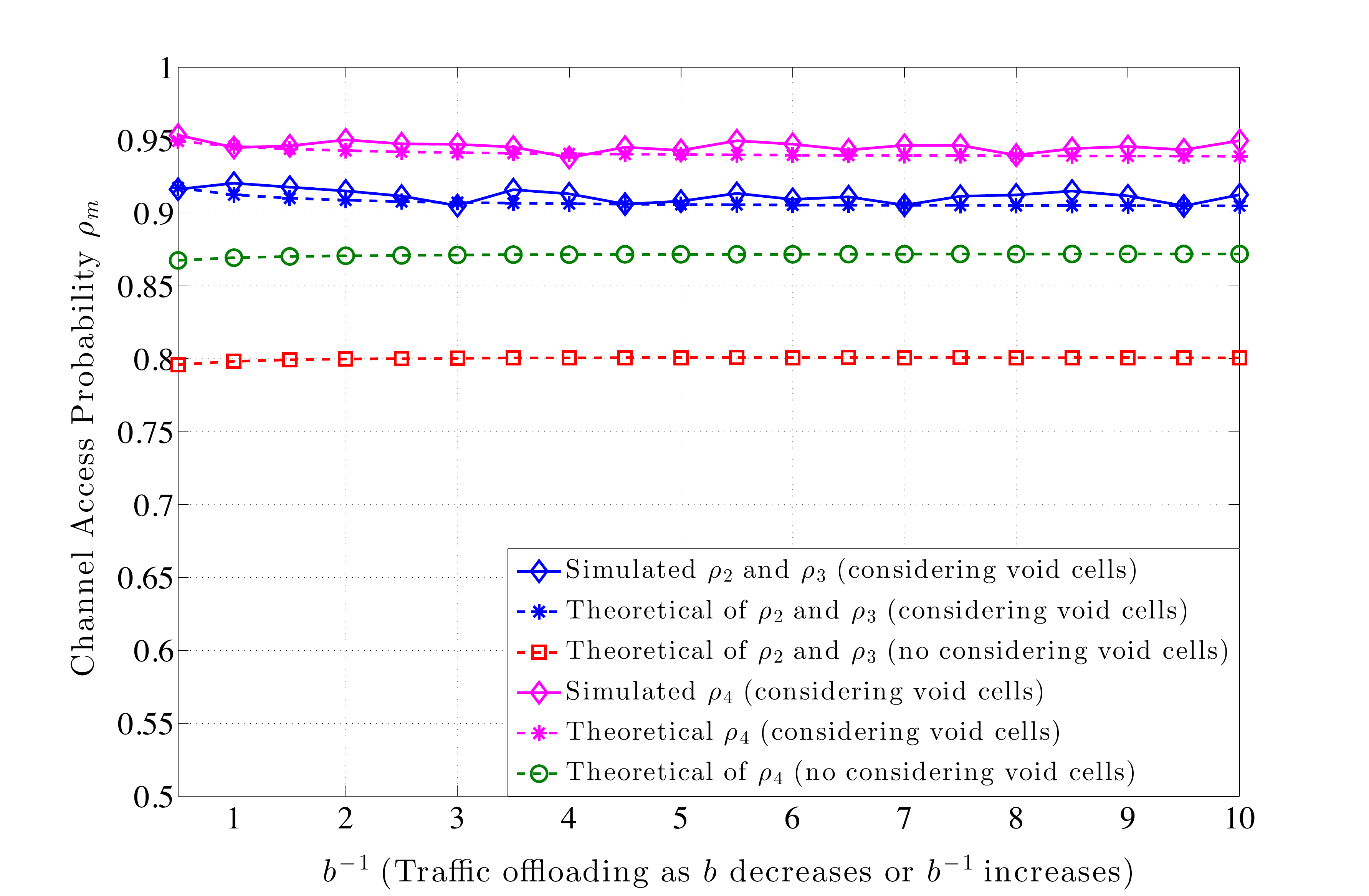}
	\caption{Simulation results of the channel access probabilities of the BSs and APs in the unlicensed spectrum when the opportunistic CSMA/CA is adopted.}
	\label{Fig:ChannelAccessProb}
\end{figure}

The void probability of an AP in a particular tier is already derived in \eqref{Eqn:VoidProb}. To verify the accuracy of \eqref{Eqn:VoidProb} and illustrate this void AP issue that cannot be overlooked in our HetNet setting here, the simulation results of the void probabilities of the BSs and APs in all tiers are presented in Fig. \ref{Fig:VoidProb}.  As shown in Fig. \ref{Fig:VoidProb}, the theoretical and simulated void probabilities are almost the same for the BSs or APs in any particular tiers, which verifies the correctness and accuracy of the derived pmf in \eqref{Eqn:CellLoadPDF} as well as the void AP probability in \eqref{Eqn:VoidProb}. In addition to the void probability of macrocell BSs, all the void probabilities of the BSs and APs in the last three tiers are actually not small at all since their intensities are not very small compared with the user intensity so that the voidness issue of the dense-deployed BSs or APs should not be carefully considered in modeling and analysis. Also, we can see that all the void probabilities of the BSs in the first three tiers increase whereas the void probability of the WiFi APs decreases while offloading traffic from the LTE network to the WiFi network. Thus, offloading or loading traffic also significantly affects the void probabilities especially for those of BSs and APs with high intensity, and this gives rise to strong impacts on the mean spectrum efficiency and per-user link throughput as shown in the following subsections. The simulation results of the channel access probabilities of the BSs and APs in all tiers are shown in Fig. \ref{Fig:ChannelAccessProb}. They not only verify the correctness of the channel access probability given in \eqref{Eqn:ChannelAssProb1} but also indicate that overlooking the void cells while modeling the contending channel behaviors among the BSs and APs can also make the channel access probabilities be seriously underestimated. Furthermore,  Fig. \ref{Fig:ChannelAccessProb} also shows that all channel access probabilities almost remain unchanged as $b^{-1}$ increases since all small BSs originally can access the unlicensed channel and the intensity of the macrocell BSs is small so that offloading traffic does not make the total intensity of BSs and APs contending the unlicensed channel alter much. 
 
\subsection{Simulation Results of Mean Spectrum Efficiency}\label{Subsec:SimMeanSpecEfff}
\begin{figure}[!h]
\centering
\includegraphics[scale=0.4]{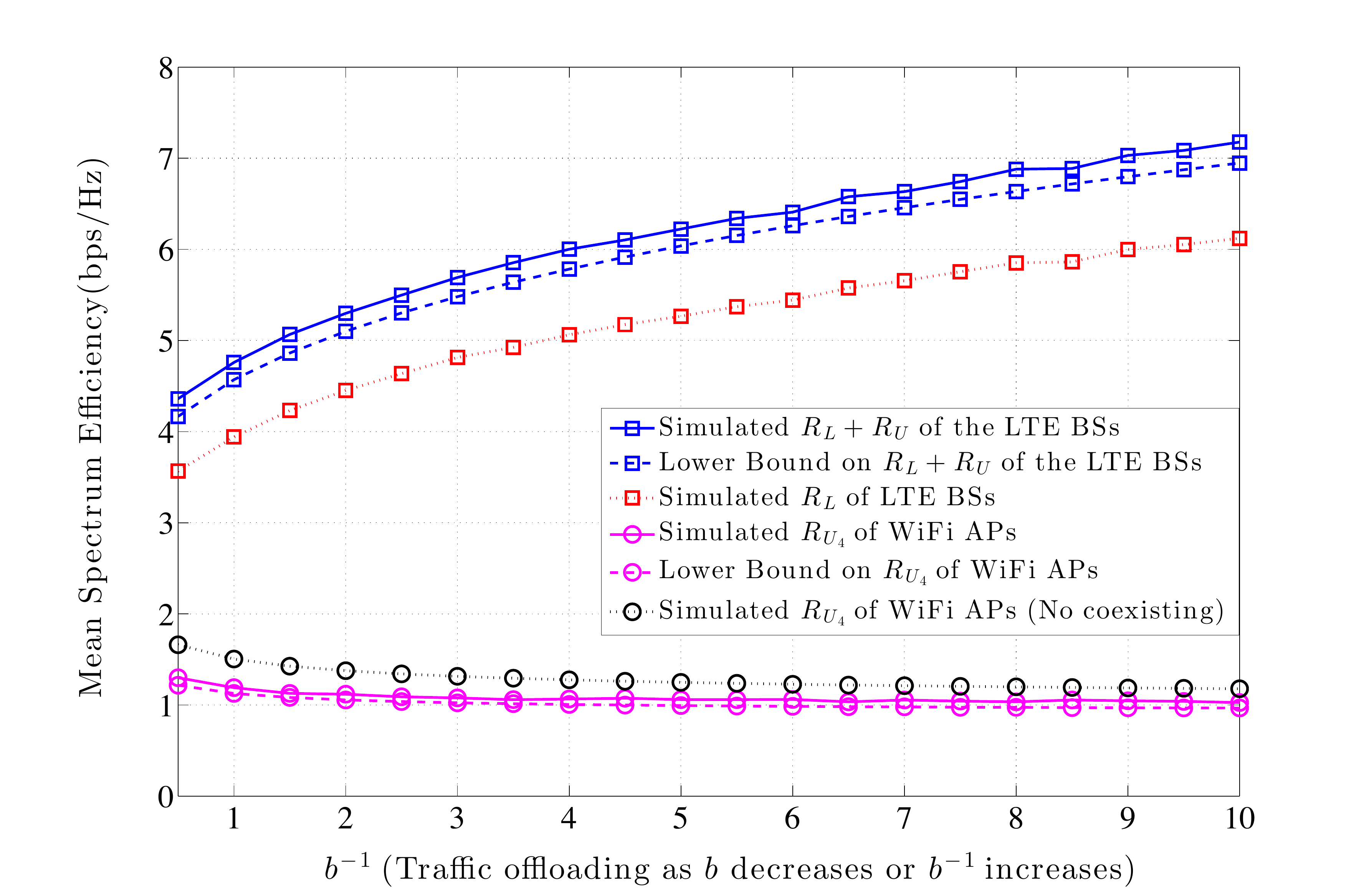}
\caption{Simulation results of the mean spectrum efficiencies in licensed and unlicensed spectra.}
\label{Fig:MeanSpectrumEff}
\end{figure}

\begin{figure}
\centering
\includegraphics[scale=0.35]{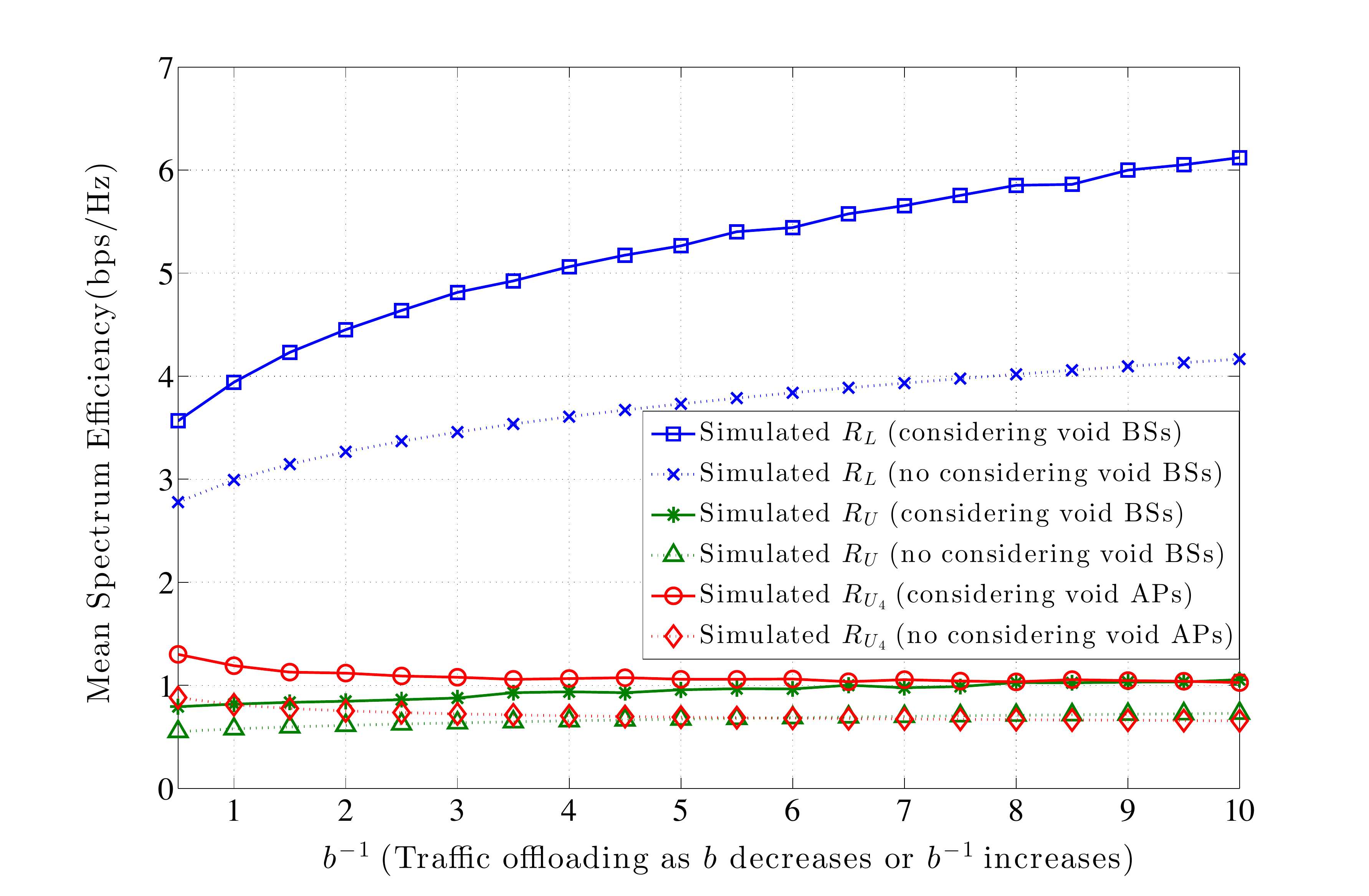}
\caption{Simulation results of the mean spectrum efficiencies with and without considering the void cell phenomenon in the interference model of the licensed and unlicensed spectra.}
\label{Fig:MeanSpectrumEffVoidComp}
\end{figure}

The simulation results of the mean spectrum efficiencies in the licensed and unlicensed spectra are shown in Fig. \ref{Fig:MeanSpectrumEff}. Since all BSs in the first three tiers use the BMSA scheme with the same bias, their mean spectrum efficiencies in the licensed spectrum are the same, i.e., $R_{\Lt_1}=R_{\Lt_2}=R_{\Lt_3}=R_{\Lt}$. Also, note that $R_{\Ut_1}=0$ since macrocell BSs do not access the unlicensed spectrum, and the BSs in the second and third tiers have the same spectrum efficiency in the unlicensed spectrum, i.e., $R_{\Ut_2}=R_{\Ut_3}=R_{\Ut}$, because they have the same channel access probability in the unlicensed spectrum. As a result, the BSs in the second and third tiers have the same sum of the mean spectrum efficiencies in the licensed and unlicensed spectra, i.e., $R_{\Lt}+R_{\Ut}$.  From Fig. \ref{Fig:MeanSpectrumEff}, we can gain a few important observations. First, the theoretical lower bound on $R_{\Lt}+R_{\Ut}$ is very tight to the simulated result of $R_{\Lt}+R_{\Ut}$, and the lower bound on $R_{\Ut_4}$ is also very close to the simulated result of $R_{\Ut_4}$. Thus, the derived lower bounds in \eqref{Eqn:UnlicensedMeanRate} and \eqref{Eqn:MeanRateLincesed2} are fairly tight, as we already emphasized this point in the previous section. Second, when LTE BSs offload their traffic, $R_{\Lt}$ significantly increases and $R_{\Ut}$ slightly increases so that $R_{\Lt}+R_{\Ut}$ significantly increases, as expected, whereas the mean spectrum efficiency $R_{\Ut_4}$ of the WiFi APs just slightly reduces. Hence, letting LTE small cell BSs and WiFi APs coexist and them share the unlicensed spectrum indeed improves their total mean spectrum efficiency in the unlicensed spectrum. Third, since the mean spectrum efficiency of the WiFi APs just slightly reduces as more traffic is offloaded from the LTE network to the WiFi network, offloading the traffic from the LTE network to the WiFi network is the best traffic management strategy for the BSs in this network setting. In Fig. \ref{Fig:MeanSpectrumEffVoidComp}, we show the simulation results of the mean spectrum efficiencies with and without considering the void AP/BS phenomenon in the interference model of the licensed and unlicensed spectra. As can be observed, $R_{\Lt}$ is significantly underestimated when the void BSs are not modeled in the interference. This validates our previous claim that the void BSs and APs should be considered in a densely deployed HetNet model. Similarly, in the unlicensed spectrum the mean spectrum efficiencies of the LTE BSs and the WiFi APs are also underestimated when void BSs and APs are not considered. As the traffic is offloaded from the LTE network to the WiFi network, the inaccuracy of the mean spectrum efficiencies of the LTE BSs without considering void BSs is exacerbated in that the void cell probabilities of the LTE BSs increase. On the contrary, the voidness impact on the mean spectrum efficiency of the WiFi network is alleviated since the offloaded traffic helps reduce the void probability of the WiFi APs.   

\subsection{Simulation Results of the Per-User Throughput and Per-User Network Throughput}\label{Subsec:SimTrafficManage}

\begin{figure}[!h]
	\centering
	\includegraphics[scale=0.35]{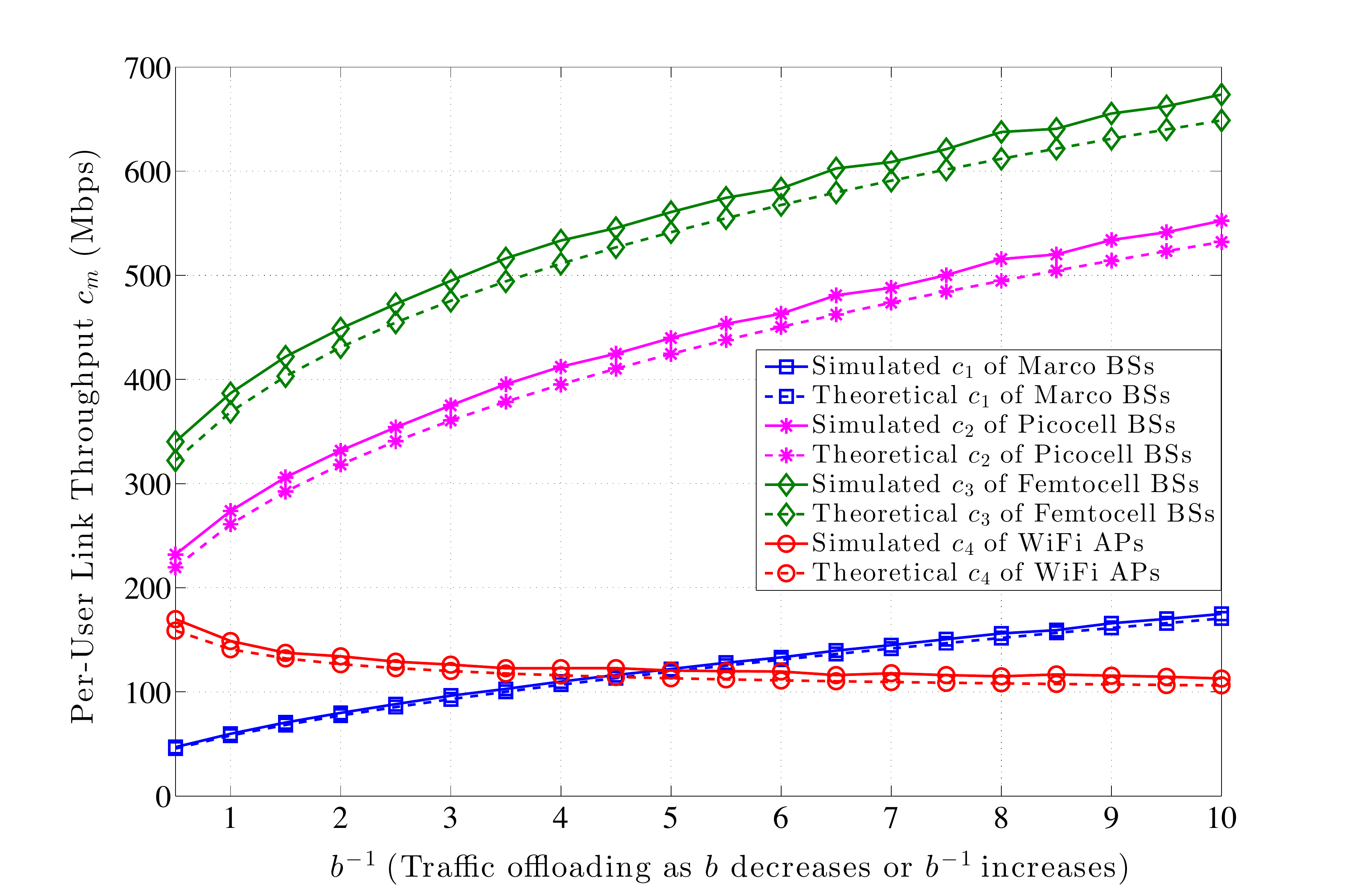}
	\caption{Simulation results of the per-user link throughputs of the BSs and APs in the four different tiers. Note that the simulated result is obtained when the decentralized traffic management scheme is performed to update/decrease $(\omega_1,\omega_2,\omega_3)$ (so as to increase $b^{-1}=P_m\mathbb{E}[H^{(s)}_m]/\omega^{\frac{\alpha}{2}}_m$.). The decentralized traffic management scheme actually updates $b^{-1}$ not much as $b^{-1}$ approaches $5$ since the per-user link throughput of the WiFi APs slowly reduces to $c_{\min}=100$ Mbps after $b^{-1}\approx 5$.}
	\label{Fig:PerUserThroughput}
\end{figure}

\begin{figure}[!h]
	\centering
	\includegraphics[scale=0.4]{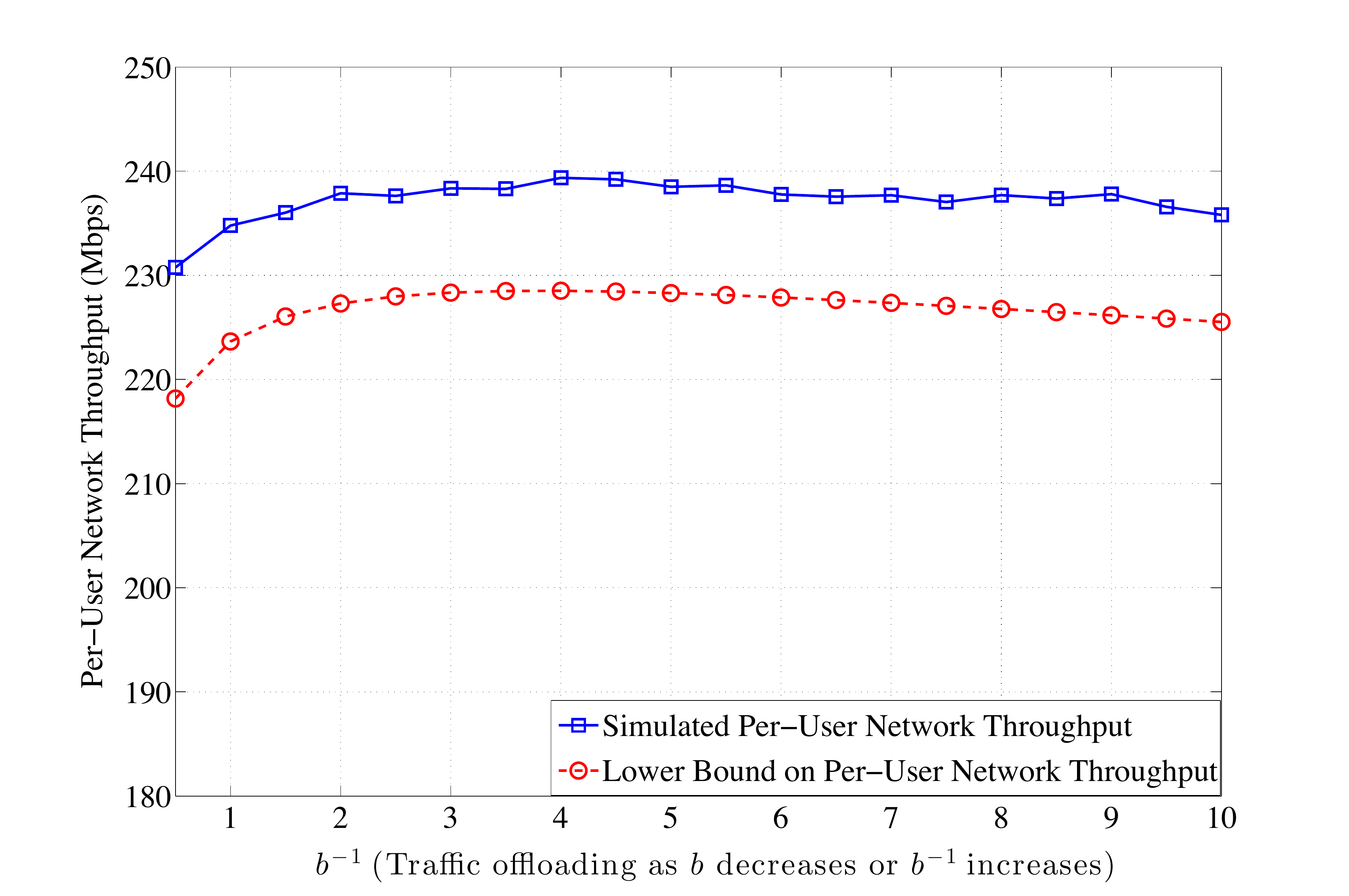}
	\caption{Simulation results of the per-user network throughputs of the BSs and APs in the four different tiers. Note that the simulated result is obtained when the centralized traffic management scheme is performed to update/decrease $(\omega_1,\omega_2,\omega_3)$ (so as to increase $b^{-1}=P_m\mathbb{E}[H^{(s)}_m]/\omega^{\frac{\alpha}{2}}_m$.). The centralized traffic management scheme actually stops to update $b^{-1}$ as $b^{-1}$ is close to $4$ since the per-user network throughput does not increase (starts to reduce) after $b^{-1}\approx 4$.}
	\label{Fig:MeanPerUserThroughput}
\end{figure}

The simulation results of the per-user link throughputs of the APs in the four different tiers are shown in Fig. \ref{Fig:PerUserThroughput} when the decentralized traffic management scheme in \eqref{Eqn:RecursiveWeightUpdate} is performed. The minimum required per-user link throughput of an WiFi AP is $c_{\min}=100$ Mbps. Initially, the unlicensed per-user link throughput of the WiFi APs is much higher than $c_{\min}$ so that all BSs start to offload their traffic as shown in Fig. \ref{Fig:PerUserThroughput}. As can be seen, offloading traffic from the LTE network to the WiFi network largely improves the per-user link throughput of the LTE BSs since $\frac{\partial c_m}{\partial \omega_m}<0$ holds in this context. Although the per-user link throughput of the WiFi APs also reduces, the throughput loss of the WiFi APs is actually not much. Accordingly, \textit{offloading the traffic from LTE to WiFi as much as possible can significantly benefit the per-user link throughput of the LTE BSs as long as $\frac{\partial c_m}{\partial \omega_m}<0$ holds and the required per-user link throughput of the WiFi APs is maintained}, as expected. However, we should notice that offloading too much traffic from the LTE network to the WiFi network could eventually give rise to the reduction in $c_m$ because $R_{\Lt_m}$ and $R_{\Ut_m}$ could both reduce in this case. Furthermore, Fig. \ref{Fig:PerUserThroughput} also illustrates that the derived lower bound on $c_m$ is a very tight to the simulated $c_m$ and its accuracy improves as the intensity of the APs is low (see the $c_4$ curve for the macrocell BSs in the figure). Finally, we provide the simulation results of the per-user network throughput of the HetNet in Fig. \ref{Fig:MeanPerUserThroughput} assuming the centralized traffic management scheme can be performed. Like the case of the per-user link throughput, offloading traffic from the LTE network to the WiFi network initially improves the per-user network throughput, but it eventually leads to the reduction in the per-user network throughput due to too much offloading. We can see the maximum of the per-user network throughput achieves at $b^{-1}\approx 4$. Hence, there indeed exists an optimal 4-tuple vector ($\omega^*_1$,$\omega^*_2$,$\omega^*_3$, $\omega^*_4$) the maximizes the per-user network throughput as shown in Lemma \ref{Lem:OptimalWeightTierm}.

\section{Conclusion}
In this paper, we consider an $M$-tier HetHet in which all APs in any particular tier form an independent PPP and the APs in the first $M-1$ tiers can simultaneously access the licensed spectrum and use the opportunistic CSMA/CA protocol to share the unlicensed spectrum with the APs in the $M$th tier. The distribution of the cell load is studied first since it can be used to find the mean cell load, the association probability of the APs in each tier. Most importantly, it characterizes the void probability of an AP in each tier that impacts the channel access probability of an AP in the unlicensed spectrum as well as the interference model especially in a dense-deployed HetNet. A novel approach is devised to find the tight lower bounds on the mean spectrum efficiencies of an AP in the licensed and unlicensed spectra for any general channel gain and AP association weight models. The per-user link throughput of an AP and the per-user network throughput are proposed and they are used to develop the decentralized and centralized traffic management schemes, respectively. These two traffic management schemes are shown to have the capability of maximizing the per-user link throughput of an AP and per-user network throughput under the constraint posed on the per-user link throughput of the APs in the $M$th tier.  

\appendix

\subsection{Proof of Theorem \ref{Thm:ShannonTransform}}\label{App:ProofShannonTransform}
The Shannon transformation of random variable $Z$ can be rewritten as
\begin{align*}
\mathcal{S}_Z(\eta)=\int_{0}^{1} \mathbb{E}\left[\frac{\eta Z}{1+y\eta Z}\right]\dif y=\int_{0}^{1} \mathbb{E}\left[\frac{1}{1/\eta Z+y}\right]\dif y.
\end{align*}
Since $\mathcal{L}_{Z^{-1}}(s)$ always exists,  for any $y\in[0,1]$ we have
\begin{align*}
\mathbb{E}\left[\frac{1}{1/\eta Z+y}\right]=\int_{0}^{\infty} e^{-uy}\mathbb{E}\left[e^{-u/\eta Z}\right]\dif u=\int_{0}^{\infty} e^{-uy}\mathcal{L}_{Z^{-1}}(u/\eta) \dif u=\int_{0}^{\infty} e^{-\eta sy}\mathcal{L}_{Z^{-1}}(s)\eta \dif s
\end{align*}
and then substituting this result into $\mathcal{S}_Z(\eta)$ yields
\begin{align*}
\mathcal{S}_Z(\eta)&=\eta\int_{0}^{\infty}\int_{0}^{1} e^{-\eta sy}\mathcal{L}_{ Z^{-1}}(s)\dif y \dif s=\int\limits_{0^+}^{\infty}\int\limits_{0}^{\infty} \frac{(1-e^{-\eta s})}{se^{\frac{s}{z}}}f_Z(z)\dif z \dif s \text{ (Letting $s=\eta z$)}.\\
&=\int_{0^+}^{\infty}\frac{(1-e^{-\eta s})}{s}\mathcal{L}_{Z^{-1}}(s) \dif s,
\end{align*}
which is exactly the result in \eqref{Eqn:IdenShannonTrans}. The result in \eqref{Eqn:IdenShannonTrans2} readily follows from \eqref{Eqn:IdenShannonTrans} and the definition of the Laplace transform of a nonnegative random variable. 

\subsection{Proof of Theorem \ref{Thm:LowBoundUnlicensedMeanRate}}\label{App:ProofLowBoundUnlicensedMeanRate}
We first show the case that all $W_{m_i}$'s are random. Let $\Delta_{\Ut_m}=P_mH_m/I_{\Ut_m}\|X_o\|^{\alpha}$ be the signal-to-interference ratio (SIR) of the typical user when AP $X_o$ is from the $m$-th tier. Thus, $R_{\Ut_m}=\mathbb{E}[\log_2(1+T_m\Xi_m\Delta_{\Ut_m})]=\xi_m\rho_m\mathbb{E}[\log_2(1+\Delta_{\Ut_m})]$ and by using the identity of the Shannon transform in Theorem \ref{Eqn:IdenShannonTrans} it can be expressed as 
\begin{align}\label{Eqn:MeanRateShannonTrans}
R_{\Ut_m} = \frac{\xi_m\rho_m}{\ln(2)}\mathbb{E}[\ln(1+\Delta_{\Ut_m})]= \frac{\xi_m\rho_m}{\ln(2)}\mathcal{S}_{\Delta_{\Ut_m}}(1) =\frac{\xi_m\rho_m}{\ln(2)}\int_{0}^{\infty}\int_{0}^{1} e^{-sy}\mathcal{L}_{\Delta^{-1}_{\Ut_m}}(s)\dif y\dif s
\end{align}
The Laplace transform of $\Delta^{-1}_{\Ut_m}$ can be further explicitly expressed as
\begin{align*}
\mathcal{L}_{\Delta^{-1}_{\Ut_m}}(s)&=\mathbb{E}\left[\exp\left(-\sum_{X_{k_i}\in\bigcup_{k=1}^M \X_k\setminus X_o}s \frac{P_kH_{k_i}V_{k_i}T_{k_i}\Xi_{k_i}\|X_o\|^{\alpha}}{P_mH_m\|X_{k_i}\|^{\alpha}}\right)\right]\\
&\stackrel{(a)}{=}\mathbb{E}\left[\exp\left(-\sum_{\widetilde{X}_{k_i}\in\bigcup_{k=1}^M \widetilde{\X}_k\setminus \widetilde{X}_o}s \frac{P_kH_{k_i}V_{k_i}T_{k_i}\Xi_{k_i}W_m\|\widetilde{X}_o\|^{\alpha}}{P_mH_mW_{k_i}\|\widetilde{X}_{k_i}\|^{\alpha}}\right)\right]\\
&\stackrel{(b)}{=} \mathbb{E}\left[\exp\left(-\sum_{\widetilde{X}_{k_i}\in\bigcup_{k=1}^M \widetilde{\X}_k\setminus \widetilde{X}_o}S_{m,k_i} \frac{\|\widetilde{X}_o\|^{\alpha}}{\|\widetilde{X}_{k_i}\|^{\alpha}}\right)\right]
\end{align*}
where $\widetilde{X}_o$, $\widetilde{X}_{m_i}$ and $\widetilde{\X}_m$ are all already defined in the Proof of Lemma \ref{Lem:AssDistPDF}, $(a)$ follows from the result in Lemma \ref{Lem:AssDistPDF} and $(b)$ is obtained by letting $S_{m,k_i}\defn \frac{sP_kH_{k_i}W_m}{P_mH_mW_{k_i}}V_{k_i}T_{k_i}\Xi_{k_i}$. Now be aware that all $T_{k_i}$'s are independent since we assume all channel gains are independent, all $V_{k_i}$'s are not completely and mutually independent since cell association could induce the location correlations between the non-void APs \cite{CHLLCW1502}\cite{CHLLCW16}, all $\Xi_{k_i}$'s are not completely and mutually independent as well since all non-void APs use CSMA/CA to access the unlicensed channel and the resulting APs accessing the channel is not a PPP any more and instead they become a Mart\'{e}n hard-core point process (MHPP) \cite{FBBBL10}\cite{CHLHCT16}.

A tractable lower bound on $\mathcal{L}_{\Delta^{-1}_{\Ut_m}}$ can be obtained by treating all $V_{k_i}$'s ($\Xi_{k_i}$'s) are independent and i.i.d. for same $k$ so that all non-void APs that have good channels and successfully access to unlicensed channel become a thinning homogeneous PPP which induces larger interference. Accordingly, we have
\begin{align*}
\mathcal{L}_{\Delta^{-1}_{\Ut_m}}(s)\geq \mathbb{E}\left[\exp\left\{-\pi\sum_{k=1}^{M}\widetilde{\lambda}_k \mathbb{E}_{S_{m,k}}\left[\int_{0}^{\infty}\left(1-e^{-S_{m,k}\left(1+\frac{x}{\|\widetilde{X}_o\|^2}\right)^{-\frac{\alpha}{2}}}\right)\dif x\right]\right\}\right]
\end{align*}
by following the proof of Proposition 2 in \cite{CHLLCW16} since $\widetilde{X}_o$ is the nearest point in the point process of $\bigcup_{m=1}^M \widetilde{X}_m$ to the typical user. Also, letting $Y$ be an exponential random variable with unit mean, i.e., $Y\sim \exp(1)$, leads to the following results:
\begin{align*}
&\int_{0}^{\infty}\left(1-e^{-S_{m,k}\left(1+\frac{x}{\|\widetilde{X}_o\|^2}\right)^{-\frac{\alpha}{2}}}\right)\dif x=\int_{0}^{\infty} \mathbb{P}\left[Y\leq S_{m,k}\left(1+\frac{x}{\|\widetilde{X}_o\|^2}\right)^{-\frac{\alpha}{2}}\right] \dif x\\
&=\int_{0}^{\infty} \mathbb{P}\left[1+\frac{x}{\|\widetilde{X}_o\|^2}\leq \left(\frac{S_{m,k}}{Y}\right)^{\frac{2}{\alpha}}\right] \dif x\\
&=\|\widetilde{X}_o\|^2\bigg(\int_{0}^{\infty} \mathbb{P}\left[u\leq \left(\frac{S_{m,k}}{Y}\right)^{\frac{2}{\alpha}}\right]\dif u-\int_{0}^{1} \mathbb{P}\left[u\leq \left(\frac{S_{m,k}}{Y}\right)^{\frac{2}{\alpha}}\right]\dif u\bigg)\\
&=\|\widetilde{X}_o\|^2\bigg[S^{\frac{2}{\alpha}}_{m,k}\Gamma\left(1-\frac{2}{\alpha}\right)+\int_{0}^{1}e^{-S_{m,k}u^{-\frac{\alpha}{2}}} \dif u -1\bigg].
\end{align*}
Hence, we have
\begin{align*}
&\mathbb{E}_{S_{m,k}}\left[\int_{0}^{\infty}\left(1-e^{-S_{m,k}\left(1+\frac{x}{\|\widetilde{X}_o\|^2}\right)^{-\frac{\alpha}{2}}}\right)\dif x\right]=\|\widetilde{X}_o\|^2(1-p_{k,0})\xi_k\rho_k\times\\
&\bigg\{\left(\frac{sP_k\bar{w}_m\bar{h}_k}{P_m\widehat{H}\bar{w}_k\bar{h}_m}\right)^{\frac{2}{\alpha}}\mathbb{E}\left[\left(\frac{H_k/\bar{h}_k}{W_k/\bar{w}_k}\right)^{\frac{2}{\alpha}}\right]\Gamma\left(1-\frac{2}{\alpha}\right)+\int_{0}^{1}\mathcal{L}_{\widehat{H}}\left(\frac{sP_k\bar{w}_m\bar{h}_ku^{-\frac{\alpha}{2}}}{P_m\bar{w}_k\bar{h}_m\widehat{H}}\right) \dif u -1\bigg\}\\
&=\|\widetilde{X}_o\|^2(1-p_{k,0})\xi_k\rho_k\ell_{\widehat{H}}\left(\frac{P_k\bar{w}_m\bar{h}_ks}{P_m\bar{w}_k\bar{h}_m\widehat{H}},\frac{2}{\alpha}\right)
\end{align*}
and this follows that
\begin{align*}
\mathcal{L}_{\Delta^{-1}_{\Ut_m}}(s)&\geq \mathbb{E}_{\widehat{H}}\left[\exp\left\{-\pi\|\widetilde{X}_o\|^2\sum_{k=1}^{M}q_{k,0}\xi_k\rho_k\widetilde{\lambda}_k\ell_{\widehat{H}}\left(\frac{sP_k\bar{w}_m\bar{h}_k}{P_m\bar{w}_k\bar{h}_m\widehat{H}},\frac{2}{\alpha}\right) \right\}\right]\\
&=\mathbb{E}_{\widehat{H}}\left[\int_{0}^{\infty}\pi\widetilde{\lambda} \exp\left\{-\pi x\widetilde{\lambda}\left(\sum_{k=1}^{M}q_{k,0}\xi_k\rho_k\vartheta_k\ell_{\widehat{H}}\left(\frac{sP_k\bar{w}_m\bar{h}_k}{P_m\bar{w}_k\bar{h}_m\widehat{H}},\frac{2}{\alpha}\right)+1\right) \right\}\dif x\right]\\
&= \mathbb{E}_{\widehat{H}}\left[\left(\sum_{k=1}^{M}q_{k,0}\xi_k\rho_k\vartheta_k\ell_{\widehat{H}}\left(\frac{sP_k\bar{w}_m\bar{h}_k}{P_m\bar{w}_k\bar{h}_m\widehat{H}},\frac{2}{\alpha}\right)+1\right)^{-1}\right].
\end{align*}

Then we know
\begin{align*}
\int_{0}^{\infty}\mathcal{L}_{\Delta^{-1}_{\Ut_m}}(s)\dif s &\geq \int_{0}^{\infty}\int_{0}^{\infty} \frac{e^{-sy}f_{\widehat{H}}(x)}{\sum_{k=1}^{M}q_{k,0}\xi_k\rho_k\vartheta_k\ell_{\widehat{H}}(\frac{sP_k\bar{w}_m\bar{h}_k}{xP_m\bar{w}_k\bar{h}_m},\frac{2}{\alpha})+1}\dif s\dif x\\
&=\int_{0}^{\infty} \frac{\int_{0}^{\infty}e^{-uxy}f_{\widehat{H}}(x)x\dif x}{\sum_{k=1}^{M}q_{k,0}\xi_k\rho_k\vartheta_k\ell_{\widehat{H}}(\frac{P_k\bar{w}_m\bar{h}_k}{P_m\bar{w}_k\bar{h}_m}u,\frac{2}{\alpha})+1}\dif u.
\end{align*}
According to \eqref{Eqn:MeanRateShannonTrans}, $R_{\Ut_m}$ can be lower bounded as
\begin{align*}
R_{\Ut_m}&=\frac{\xi_m\rho_m}{\ln(2)}\int_{0}^{\infty}\int_{0}^{1} e^{-sy}\mathcal{L}_{\Delta^{-1}_{\Ut_m}}(s)\dif y\dif s\geq \frac{\xi_m\rho_m}{\ln (2)} \int_{0}^{\infty} \frac{\int_{0}^{\infty}(\int_{0}^{1}e^{-uxy}\dif y)f_{\widehat{H}}(x)x\dif x}{\sum_{k=1}^{M}q_{k,0}\xi_k\rho_k\vartheta_k\ell_{\widehat{H}}(\frac{P_k\bar{w}_m\bar{h}_k}{P_m\bar{w}_k\bar{h}_m}u,\frac{2}{\alpha})+1}\dif u,
\end{align*}
which is exactly the lower bound in \eqref{Eqn:UnlicensedMeanRate} by carrying out those inner double integrals for variables $x$ and $y$. For the case that all $W_{m_i}$'s are constant $\bar{w}_m$ for all $m\in\mathcal{M}$, letting all $W_m$ leave in function $\ell(\cdot,\cdot)$ yields the result in \eqref{Eqn:UnlicensedMeanRate2} with $\widehat{H}=H_m/\bar{h}_m$ for all $m\in\mathcal{M}$.

\subsection{Proof of Theorem \ref{Thm:DecenTrafficMang}}\label{App:ProofDecenTrafficMang}
First notice that the coefficient of $\omega_m(n)$ in \eqref{Eqn:RecursiveWeightUpdate} has the physical meaning that $c^*(n)$ is normalized by the per-user link throughput of a tier-$m$ AP in the unlicensed spectrum. The tier-$m$ APs will offload (load) its traffic in the next time if this coefficient at time $n$ is smaller (greater) than one.  In other words, the per-user link throughput of the tier-$m$ APs can continuously increase (decrease) under the traffic management scheme in \eqref{Eqn:RecursiveWeightUpdate} as long as it is smaller (larger) than $c^*(n)$ for all $n\in\mathbb{N}$. Now we want to show that $\omega_m(n)$ converges to a steady-state value $\omega^*_m$ as $n$ goes to infinity. Consider a Lyapunov function $V(n)\defn (\omega_m(n)-\omega^*_m/2)^2$ and we can have the following result
\begin{align*}
V(n+1)-V(n)&=(\omega_m(n+1)-\omega^*_m/2)^2-(\omega_m(n)-\omega^*_m/2)^2\\
&=\left[\omega_m(n+1)-\omega_m(n)\right]\left[\omega_m(n+1)+\omega_m(n)-\omega^*_m\right]\\
&=
\left[\frac{c^*(n)N_m(n)}{C_m(n)}-1\right]\left[\frac{c^*(n)N_m(n)}{C_m(n)}+1-\frac{\omega_m^*}{\omega_m(n)}\right] \omega^2_m(n) <0,
\end{align*} 
which yields the following two constraints
\begin{align*}
1< \frac{c^*(n)N_m(n)}{C_m(n)}<\frac{\omega_m^*}{\omega_m(n)}-1\,\,\text{  and  }\,\,
\frac{\omega_m^*}{\omega_m(n)}-1<\frac{c^*(n)N_m(n)}{C_m(n)}<1.
\end{align*}
As long as $w_m(n)$ and $\frac{c^*(n)N_m(n)}{C_m(n)}$ satisfy these two constraints, $\omega_m(n)$ converges to $\omega^*_m$ as $n$ goes to infinity based on the Foster-Lyapunov criterion \cite{SPMRLT93}. In other words, as $n$ goes to infinity, $c^*(n)$ will converge to $\sup_n\{C_m(n)/N_m(n)\}$. Also, these two constraints always hold since $\frac{c^*(n)N_m(n)}{C_m(n)}<1$ (i.e. $C_m/N_m(n)>c^*(n)$) makes $\omega_m(n)$ reduce and approach to $\omega^*_m$ if $\omega_m(n)>\omega^*_m/2$ and $\frac{c^*(n)N_m(n)}{C_m(n)}>1$ (i.e. $C_m/N_m(n)>c^*(n)$) makes $\omega_m(n)$ reduce and approach to $\omega^*_m$ if $\omega_m(n)<\omega^*_m/2$. Therefore, the decentralized traffic management scheme in \eqref{Eqn:RecursiveWeightUpdate} will make $\omega_m(n)$ converge to  $\omega^*_m$ as time goes to infinity and thus we must have $0<\frac{c^*(n)N_m(n)}{C_m(n)}<1$, i.e., $C_m(\omega^*_m)/N_m(\infty)> c^*(\infty)$ for all $m\in\mathcal{M}$, and $c_M\geq c_{\min}$ is surely satisfied. In addition, since $N_m(\infty)=\mu\vartheta_m(\omega^*_m)/\lambda_m$, in the steady state $\omega^*_m$ must satisfy the following constraint
\begin{align*}
\frac{\mu c^*\vartheta_m(\omega^*_m)}{\lambda_mC_m(\omega^*_m)}=1 \Rightarrow \omega^*_m = \frac{C_m(\omega^*_m)}{\mu c^*}\left(\sum_{k\in\mathcal{M}\setminus m}\lambda_k\omega_k+\lambda_m\omega^*_m\right),
\end{align*}
which indicates $\omega^*_m$ is the fixed point of $\Upsilon_m(x)$ in \eqref{Eqn:FunOptWeight}.




\bibliographystyle{ieeetran}
\bibliography{IEEEabrv,Ref_SpectrumSharingHetNets}

\begin{thebibliography}{10}
\providecommand{\url}[1]{#1}
\csname url@samestyle\endcsname
\providecommand{\newblock}{\relax}
\providecommand{\bibinfo}[2]{#2}
\providecommand{\BIBentrySTDinterwordspacing}{\spaceskip=0pt\relax}
\providecommand{\BIBentryALTinterwordstretchfactor}{4}
\providecommand{\BIBentryALTinterwordspacing}{\spaceskip=\fontdimen2\font plus
\BIBentryALTinterwordstretchfactor\fontdimen3\font minus
  \fontdimen4\font\relax}
\providecommand{\BIBforeignlanguage}[2]{{%
\expandafter\ifx\csname l@#1\endcsname\relax
\typeout{** WARNING: IEEEtran.bst: No hyphenation pattern has been}%
\typeout{** loaded for the language `#1'. Using the pattern for}%
\typeout{** the default language instead.}%
\else
\language=\csname l@#1\endcsname
\fi
#2}}
\providecommand{\BIBdecl}{\relax}
\BIBdecl

\bibitem{MBMSACO13}
M.~Bennis, M.~Simsek, A.~Czylwik \emph{et~al.}, ``When cellular meets {WiFi} in
  wireless small cell networks,'' \emph{{IEEE} Commun. Mag.}, vol.~51, no.~6,
  pp. 44--50, Jun. 2013.

\bibitem{HZXCWGSW15}
H.~Zhang, X.~Chu, W.~Guo, and S.~Wang, ``Coexistence of {Wi-Fi} and
  heterogeneous small cell networks sharing unlicensed spectrum,'' \emph{{IEEE}
  Commun. Mag.}, vol.~53, no.~3, pp. 158--164, Mar. 2015.

\bibitem{RZMWLXCZZXSLLX15}
R.~Zhang, M.~Wang, L.~X. Cai, Z.~Zheng, X.~Shen, and L.-L. Xie,
  ``{LTE}-unlicensed: the future of spectrum aggregation for cellular
  networks,'' \emph{{IEEE} Wireless Commun. Mag.}, vol.~22, no.~3, pp.
  150--159, Jun. 2015.

\bibitem{REAPDT07}
R.~Etkin, A.~Parekh, and D.~Tse, ``Spectrum sharing for unlicensed bands,''
  \emph{{IEEE} J. Select. Areas Commun.}, vol.~25, no.~3, pp. 517 -- 528, Apr.
  2007.

\bibitem{HYPPHCNRP07}
H.~Yomo, P.~Popovski, H.~C. Nguyen, and R.~Prasad, ``Adaptive frequency rolling
  for coexistence in the unlicensed band,'' \emph{{IEEE} Trans. Wireless
  Commun.}, vol.~6, no.~2, pp. 598--608, Oct. 2007.

\bibitem{YMSKAH09}
Y.~M. Shobowale and K.~A. Hamdi, ``A unified model for interference analysis in
  unlicensed frequency bands,'' \emph{{IEEE} Trans. Wireless Commun.}, vol.~8,
  no.~8, pp. 4004--4013, Aug. 2009.

\bibitem{JBENNJR12}
J.~B. Ernst, N.~Nasser, and J.~Rodrigues, ``Co-channel interference modelling
  between {RATs} in heterogeneous wireless networks,'' in \emph{IEEE Int. Conf.
  on Commun.}, Jun. 2012, pp. 5321--5325.

\bibitem{JJQLHNAPGW14}
J.~Jeon, Q.~Li, H.~Niu, A.~Papathanassiou, and G.~Wu, ``{LTE} in the unlicensed
  spectrum: A novel coexistence analysis with {WLAN} systems,'' in \emph{IEEE
  Global Commun. Conf.}, Dec. 2014, pp. 3459--3464.

\bibitem{CHLHCT16}
\BIBentryALTinterwordspacing
C.-H. Liu and H.-C. Tsai, ``On the limits of coexisting coverage and capacity
  in multi-{RAT} heterogeneous networks,'' \emph{submitted to IEEE Trans.
  Wireless Commun.}, 2016. [Online]. Available:
  \url{http://arxiv.org/abs/1602.02250}
\BIBentrySTDinterwordspacing

\bibitem{XDCHLLCWXZ16}
X.~Ding, C.-H. Liu, L.-C. Wang, and X.~Zhao, ``Coexisting success and
  throughput of multi-rat wireless networks with unlicensed band access,''
  \emph{IEEE Wireless Comm. Letters}, vol.~5, no.~1, pp. 4--7, Feb. 2016.

\bibitem{HCTCHL16}
H.-C. Tsai, C.-H. Liu, and L.-C. Wang, ``An analytical approach to coexisting
  evaluation in {Multi-RAT} heterogeneous networks with opportunistic
  {CSMA/CA},'' in \emph{IEEE Int. Conf. on Commun.}, May 2016, pp. 1--6.

\bibitem{QCGYHSAMGYLAH16}
Q.~Chen, G.~Yu, H.~Shan, A.~Maaref, G.~Y. Li, and A.~Huang, ``Cellular meets
  {WiFi}: Traffic offloading or resource sharing?'' \emph{{IEEE} Trans.
  Wireless Commun.}, vol.~15, no.~5, pp. 3354--3367, May 2016.

\bibitem{AREWPCAIZD15}
A.~R. Elsherif, W.-P. Chen, A.~Ito, and Z.~Ding, ``Resource allocation and
  inter-cell interference management for dual-access small cells,''
  \emph{{IEEE} J. Select. Areas Commun.}, vol.~33, no.~6, pp. 1082--1096, Jun.
  2015.

\bibitem{FLEBEEMCBRY15}
F.~Liu, E.~Bala, E.~Erkip, M.~C. Beluri, and R.~Yang, ``Small-cell traffic
  balancing over licensed and unlicensed bands,'' \emph{{IEEE} Trans. Veh.
  Technol.}, vol.~64, no.~12, pp. 5850--5865, Dec. 2015.

\bibitem{ABCIPZ14}
A.~Bhorkar, C.~Ibars, and P.~Zong, ``Performance analysis of {LTE} and {WiFi}
  in unlicensed band using stochastic geometry,'' in \emph{IEEE Globecom
  Workshop on Heterogeneous and Small Cell Networks}, Sep. 2014, pp.
  1310--1314.

\bibitem{SSISDR15}
S.~Sagari, I.~Seskar, and D.~Raychaudhuri, ``Modeling the coexistence of {LTE}
  and {WiFi} heterogeneous networks in dense deployment scenarios,'' in
  \emph{IEEE ICC workshop on LTE in unlicensed bands: potentials and
  challenges}, Jun. 2015, pp. 2301--2306.

\bibitem{CHLLCW1602}
C.-H. Liu and L.-C. Wang, ``Modeling and analysis of coexisting multiple radio
  access technologies in heterogeneous wireless networks,'' in \emph{IEEE Int.
  Conf. on Computing, Networking and Commun.}, Feb. 2016, pp. 1--5.

\bibitem{YLFBJGA16}
Y.~Li, F.~Baccelli, J.~G. Andrews, T.~D. Novlan, and J.~C. Zhang, ``Modeling
  and analyzing the coexistence of {Wi-Fi} and {LTE} in unlicensed spectrum,''
  \emph{{IEEE} Trans. Wireless Commun.}, vol.~15, no.~9, pp. 6310--6326, Dec.
  2016.

\bibitem{XWTQMSJL17}
X.~Wang, T.~Q.~S. Quek, M.~Sheng, and J.~Li, ``Throughput and fairness analysis
  of {Wi-Fi} and {LTE-U} in unlicensed band,'' \emph{{IEEE} J. Select. Areas
  Commun.}, vol.~35, no.~PP, accepted and to appear in 2017.

\bibitem{3GPPLAA15}
{3GPP TR 36.889 v13}, ``Study on licensed-assisted access to unlicensed
  spectrum,'' May 2015.

\bibitem{DSWKJM96}
D.~Stoyan, W.~Kendall, and J.~Mecke, \emph{Stochastic Geometry and Its
  Applications}, 2nd~ed.\hskip 1em plus 0.5em minus 0.4em\relax New York: John
  Wiley and Sons, Inc., 1996.

\bibitem{CHLLCW1502}
C.-H. Liu and L.-C. Wang, ``Random cell association and void probability in
  poisson-distributed cellular networks,'' in \emph{IEEE Int. Conf. on
  Commun.}, Jun. 2015, pp. 2816--2821.

\bibitem{CHLLCW16}
------, ``Optimal cell load and throughput in green small cell networks with
  generalized cell association,'' \emph{{IEEE} J. Select. Areas Commun.},
  vol.~34, no.~5, pp. 1058--1072, May 2016.

\bibitem{HSJYJSXPJGA12}
H.-S. Jo, Y.~J. Sang, X.~Ping, and J.~G. Andrews, ``Heterogeneous cellular
  networks with flexible cell association: A comprehensive downlink {SINR}
  analysis,'' \emph{{IEEE} Trans. Wireless Commun.}, vol.~11, no.~10, pp.
  3484--3495, Oct. 2012.

\bibitem{CHLKLF16}
C.-H. Liu and K.~L. Fong, ``Fundamentals of the downlink green coverage and
  energy efﬁciency in heterogeneous networks,'' \emph{{IEEE} J. Select. Areas
  Commun.}, vol.~34, no.~12, pp. 3271--3287, Dec. 2016.

\bibitem{JGAFBRKG11}
J.~G. Andrews, F.~Baccelli, and R.~K. Ganti, ``A tractable approach to coverage
  and rate in cellular networks,'' \emph{{IEEE} Trans. Commun.}, vol.~59,
  no.~11, pp. 3122--3134, 2011.

\bibitem{PXCHLJGA13}
P.~Xia, C.-H. Liu, and J.~G. Andrews, ``Downlink coordinated multi-point with
  overhead modeling in heterogeneous cellular networks,'' \emph{{IEEE} Trans.
  Wireless Commun.}, vol.~12, no.~8, pp. 4025--4037, Jun. 2013.

\bibitem{DPB99}
D.~P. Bertsekas, \emph{Nonlinear Programming}, 2nd~ed.\hskip 1em plus 0.5em
  minus 0.4em\relax Belmont, MA: Athena Scientific, 1999.

\bibitem{FBBBL10}
F.~Baccelli and B.~B{\l}aszczyszyn, ``Stochastic geometry and wireless
  networks: Volume {II A}pplications,'' \emph{Foundations and Trends in
  Networking}, vol.~3, no. 3-4, pp. 249--449, 2010.

\bibitem{SPMRLT93}
S.~P. Meyn and R.~L. Tweedie, \emph{Markov Chains and Stochastic
  Stability}.\hskip 1em plus 0.5em minus 0.4em\relax New York: Springer-Verlag,
  1993.

\end{thebibliography}

\end{document}